\RequirePackage{fix-cm}
\documentclass[epjc3]{svjour3}%one column
\smartqed  % flush right qed marks, e.g. at end of proof
\RequirePackage{graphicx}
\journalname{Eur. Phys. J. C}
\usepackage{stackrel}
\usepackage{amscd}
\usepackage{amsbsy}
\usepackage{array}

\usepackage{amssymb, amsmath}%aggiunto dopo
\usepackage{amsmath,bm,cite,color}
\usepackage{tikz}
\usepackage{graphicx}
\usepackage{caption}
\usepackage{subcaption}
\usepackage{lineno}
\newcommand{\mathsym}[1]{{}}

\usepackage{epsfig}
\usepackage{epsf}
\usepackage{rotating}
\usepackage{color}
\usepackage{hyperref}
\usepackage[labelformat=parens,labelsep=quad,skip=3pt]{caption}
\usepackage{cancel}
\newfont{\tenmsb}{msbm10 scaled\magstep1}

\let\ssection=\section\renewcommand{\section}
{\setcounter{equation}{0}\ssection}

\newcommand{\half}{{\scriptstyle{\frac{1}{2}}}}

\newcommand{\cO}{{\mathcal O}}
\newcommand{\lf}{\left (}
\newcommand{\lfq}{\left [}

\newcommand{\rg}{\right )}
\newcommand{\rgq}{\right ]}

\def\smallover#1/#2{\hbox{$\textstyle{#1\over#2}$}}
\def\beq{\begin{equation}}
\def\eeq{\end{equation}}
\def\beq{\begin{equation}}
\def\eeq{\end{equation}}
\def\bea{\begin{eqnarray}}
\def\eea{\end{eqnarray}}
\newcommand{\nn}{\nonumber}
\def\lf{\left(}
\def\rg{\right)}
\def\lq{\left[}
\def\rq{\right]}
\def\lgr{\left\{}
\def\rgr{\right\}}

\def\vr{{\vec r}}

\def\*{{\star}}

\def\hO{{\hat \Omega}}

\def\nima{{Nucl. Instr. \& Meth. in Physics Res. A }}

 \usepackage[final]{changes}

\definechangesauthor[color=green]{vito}
\definechangesauthor[color=purple]{3}
\definechangesauthor[color=blue]{2}
\definechangesauthor[color=red]{1}

\begin{document}

%% \title{Coded masks  for particle physics  events%Insert your title here%\thanksref{t1}
\title{Coded masks for imaging of neutrino events%Insert your title here%\thanksref{t1}
}


\author{ M.~Andreotti~\thanksref{addr4,addr5}\and	           P.~Bernardini~\thanksref{addr1,addr2}\and 
              A.~Bersani~\thanksref{addr6}\and 	                   S.~Bertolucci~\thanksref{BO1,BO2}\and
              S.~Biagi~\thanksref{LNS}\and
              A.~Branca~\thanksref{addr8,addr9}\and                 C.~Brizzolari~\thanksref{addr8,addr9}\and            
              G.~Brunetti~\thanksref{addr8,addr9}\and                I.~Cagnoli~\thanksref{BO1,BO2}\and
              R.~Calabrese~\thanksref{addr4,addr5}\and            A.~Caminata~\thanksref{addr6}\and 
              A.~Campani~\thanksref{addr6,addr7}\and              P.~Carniti~\thanksref{addr8,addr9}\and 
 	      R.~Cataldo~\thanksref{addr1}\and                          C.~Cattadori~\thanksref{addr9}\and 
              S.~Cherubini~\thanksref{LNS}\and                          V.~Cicero~\thanksref{BO1,BO2}\and
              M.~Citterio~\thanksref{addr11}\and                         S.~Copello~\thanksref{addr6,addr7}\and
              P.~Cova~\thanksref{addr11,addr12}\and                 E.~Cristaldo~Morales~\thanksref{addr8,addr9}\and 
              S.~Davini~\thanksref{addr6}\and                             N.~Delmonte~\thanksref{addr11,addr12}\and
              G.~De~Matteis~\thanksref{addr1,addr2,addr3}\and 
              S.~Di~Domizio~\thanksref{addr6,addr7}\and          L.~Di~Noto~\thanksref{addr6,addr7}\and 
              C.~Distefano~\thanksref{LNS}\and
              T.~Giammaria~\thanksref{addr4,addr5}\and            M.~Guarise~\thanksref{addr4,addr5}\and 
              A.~Falcone~\thanksref{addr8,addr9}\and                F.~Ferraro~\thanksref{addr6,daga1}\and  %% a Milano dal 1/2/21
              M.~Fiorini~\thanksref{addr4,addr5}\and                  N.~Gallice~\thanksref{addr10,addr11}\and              
              C.~Gotti~\thanksref{addr9}\and                               M.~Guerzoni~\thanksref{BO2}\and              
              M.A.~Iliescu~\thanksref{e1,LNF}\and                      G.~Ingratta~\thanksref{BO1,BO2}\and
              M.~Lazzaroni~\thanksref{addr10,addr11}\and         I.~Lax~\thanksref{BO2}\and   
              G.~Laurenti~\thanksref{BO2}\and
              A.~Leaci~\thanksref{addr1,addr2}\and                    E.~Luppi~\thanksref{addr4,addr5}\and
              L.~Martina~\thanksref{e2,addr1,addr2}\and            N.~Mauri~\thanksref{BO1,BO2}\and              
              A.~Minotti~\thanksref{addr4,addr5,daga2}\and       N.~Moggi~\thanksref{BO1,BO2}\and
              E.~Montagna~\thanksref{BO1,BO2}\and                A.~Montanari~\thanksref{BO2}\and
              D.~Montanino~\thanksref{addr1,addr2}\and           M.~Pallavicini~\thanksref{addr6,addr7}\and            
              M.~Panareo~\thanksref{addr1,addr2}\and              E.G.~Parozzi~\thanksref{addr8,addr9}\and 
              L.~Pasqualini~\thanksref{BO1,BO2}\and                L.~Patrizii~\thanksref{BO2}\and
              G.~Pessina~\thanksref{addr9}\and                          F.~Poppi~\thanksref{BO1,BO2}\and 
              M.~Pozzato~\thanksref{BO2}\and                           V.~Pia~\thanksref{BO1,BO2}\and
              S.~Riboldi~\thanksref{addr10,addr11}\and              G.~Riccobene~\thanksref{LNS}\and
              P.~Sala~\thanksref{addr11}\and                              P.~Sapienza~\thanksref{LNS}\and
              F.~Schifano~\thanksref{addr4,addr5}\and               G.~Sirri~\thanksref{BO2}\and              
              M.~Spanu~\thanksref{addr8,addr9}\and                  L.~Stanco~\thanksref{addr13}\and 
              A.~Surdo~\thanksref{addr2}\and                              A.~Taibi~\thanksref{addr4,addr5}\and
              M.~Tenti~\thanksref{BO2}\and                                 F.~Terranova~\thanksref{addr8,addr9}\and 
              G.~Testera~\thanksref{addr6}\and                           L.~Tomassetti~\thanksref{addr4,addr5}\and
              M.~Torti~\thanksref{addr8,addr9}\and                     N.~Tosi~\thanksref{BO2}\and                          
              R.~Travaglini~\thanksref{BO2}\and                          L.~Uboldi~\thanksref{addr10,addr11,daga3}\and
              M.~Vicenzi~\thanksref{addr6,addr7}\and                A.~Zani~\thanksref{addr11}\and
              S.~Zucchelli~\thanksref{BO1,BO2}\\                       (NU@FNAL Collaboration)
}

%\thankstext{e1}{e-mail: fauthor@example.com}
\thankstext{e1}{ e-mail: mihai.iliescu@lnf.infn.it}     %% ORCID		0000-0002-9648-3451
\thankstext{e2}{ e-mail: luigi.martina@le.infn.it}      %% ORICD		0000-0003-0950-1365
\thankstext{daga1}{ Presently at Dipartimento di Fisica, Universit\`a di Milano}
\thankstext{daga2}{ Presently at Dipartimento di Fisica, Universit\`a di Milano-Bicocca}
\thankstext{daga3}{ Presently at CERN}


\institute{        Dipartimento di Fisica e Scienze della Terra, Universit\`a di Ferrara, via G. Saragat 1, 44122 Ferrara, Italy\label{addr4}
              \and Istituto Nazionale di Fisica Nucleare, Sezione di Ferrara, via G. Saragat 1, 44122 Ferrara, Italy\label{addr5}
              \and Dipartimento di Matematica e Fisica "Ennio De Giorgi", Universit\`a del Salento, via per Arnesano, 73100 Lecce, Italy\label{addr1}
              \and Istituto Nazionale di Fisica Nucleare, Sezione di Lecce, via per Arnesano, 73100 Lecce, Italy\label{addr2}
              \and Dipartimento di Fisica, Universit\`a di Genova, via Dodecaneso 33, 16146 Genova, Italy\label{addr6}
              \and Dipartimento di Fisica, Universit\`a di Bologna, viale C. Berti Pichat 6/2, 40127 Bologna, Italy\label{BO1}
              \and Istituto Nazionale di Fisica Nucleare, Sezione di Bologna, viale C. Berti Pichat 6/2, 40127 Bologna, Italy\label{BO2}   
              \and Istituto Nazionale di Fisica Nucleare, Laboratori Nazionali del Sud, via S. Sofia 62, 95125 Catania, Italy\label{LNS}
              \and Universit\`a di Milano-Bicocca, piazza della Scienza 3, 20126 Milano, Italy\label{addr8}
              \and Istituto Nazionale Fisica Nucleare, Sezione di Milano-Bicocca, piazza della Scienza 3, 20126 Milano, Italy\label{addr9}
              \and Istituto Nazionale Fisica Nucleare, Sezione di Genova, via Dodecaneso 33, 16146 Genova, Italy\label{addr7}
              \and Istituto Nazionale Fisica Nucleare, Sezione di Milano, via Celoria 16, 20133 Milano, Italy\label{addr11}
              \and Dipartimento di Ingegneria, Universit\`a di Parma, parco Area delle Scienze 181/A, 43124 Parma, Italy\label{addr12}
              \and GNFM-INDAM, Citt\`a Universitaria, piazzale Aldo Moro 5, 00185 Roma, Italy\label{addr3}
              \and Dipartimento di Fisica, Universit\`a di Milano, via Celoria 16, 20133 Milano, Italy\label{addr10}
              \and Istituto Nazionale Fisica Nucleare, Laboratori Nazionale di Frascati, via E. Fermi 54, 00044 Frascati, Italy\label{LNF}
              \and Istituto Nazionale Fisica Nucleare, Sezione di Padova, Padova, Italy\label{addr13}
               }

\date{Received: date / Accepted: date}

\maketitle

\begin{abstract}   

\noindent The capture of scintillation light emitted by liquid Argon and Xenon under molecular excitations by charged particles is still a challenging task.
Here we present a first attempt to design a device able to have a sufficiently high photon detection efficiency, in order to reconstruct the path of ionizing particles.
The study is based on the use of masks to encode the light signal combined with single-photon detectors, showing the capability to detect tracks over focal distances of about tens of centimeters. From numerical simulations it emerges that it is possible to successfully decode 
and recognize signals, even of rather complex topology, with a relatively limited number of acquisition channels.
Thus, the main aim  is to elucidate a proof of principle of a  technology developed in very different contexts, but which has potential applications in liquid argon detectors that require a fast reading.
The findings support us to think that such innovative technique could be very fruitful in a new 
generation of detectors devoted to neutrino physics. 
\end{abstract}

\newpage

\section{Introduction}\label{sec:intro}

This work is aimed at introducing a new and more efficient  collection  method  of prompt photons emitted by charged particle in noble liquid filling Time Projection Chambers (TPCs), in order to obtain   track images, instead of simple triggering signals.  
As it is known, noble elements in the liquid phase (LAr, LXe) are used as target and detector in high energy physics. In these liquid gases, relativistic charged particles produce large 
amount of scintillation light in the Vacuum UltraViolet (VUV) range. However, in TPCs the event reconstruction is just based on the collection 
of drift electrons and the fast light signal is exploited only to set the trigger time $t_0$ for the data acquisition. The benefits of this novel technique are several, as rate capability, especially relevant for 
accelerator based experiments, and possibility to work in magnetic field. On the other side, such an imaging detector presents also critical issues. For example, performance 
of conventional optics in VUV range is very poor and readout electronics must be operated in cryogenic conditions with single-photon detection capability.

In order to face these new challenges we are conceiving a system where the light signal is filtered by \emph{Coded Masks} and read by \emph{Silicon Photomultipliers} (SiPMs).
The latters guarantee the required performances and offer the advantage of robustness, large number of densely packed small pixels and strong reduction of dark noise
    at low temperature. The coded masks should have a sufficiently high photon detection efficiency without the use of special materials and complex designs. They provide a sufficiently wide and deep 
field of view and a large aperture, in such a way to minimize the number of SiPMs.

As proof of principle of the above quoted imaging method, we assume the Near Detector~\cite{duneND} of the DUNE experiment~\cite{dune} as an inspiring situation, without entering into a full realistic modeling of it.
Taking into account that the detector will be hit by the most intense high-energy neutrino beam, the high-rate capability is mandatory. In particular we plan to have a 
LAr volume in the SAND apparatus (System for on-Axis Neutrino Detection) where a $0.6\ T$ magnetic field is present, equipped with coded masks and SiPM arrays. 
The typical energies of the particles (mainly muons) produced in neutrino interactions are sufficiently high to generate  $\approx 10^4\ photons/sr/cm$ at the  
$\lambda_{VUV} \simeq 128\, nm$ wavelength. The imaging reconstruction of neutrino events in the LAr target will be exploited not only to continuously monitor the neutrino-beam 
spectrum but also to measure neutrino fluxes and cross-section in LAr in order to constrain nuclear effects.
However, for the specific application described above, the designers will face the problem of balancing the beam spill length and the long time scintillation constant, leading to possible multiple interactions. This aspect is outside the scope of the present paper and it will concern an evaluation of: the effective probability of events occurring in the fiducial volume "seen" by the masks, the electronic timing, the filtering of the diffused light background and the optical  effects of mixture of noble liquids on the light transmission.

In this work, the basic principles of the imaging technique with coded masks are presented. The exploitation of other optical schemes for VUV photons is also under consideration and 
they will be the topic of future papers.

\section{ Imaging by Coded Masks}
It is well understood that a small pinhole is required to achieve high spatial resolution. But a single pinhole also dims the light in the image, so much that it may be below the sensitivity 
of light-sensors. A matrix of multiple pinholes increases light collection, but the source reconstruction from multiple superimposed images becomes more convoluted, and this approach 
requires to exploit fast numerical methods~\cite{Mertz1,Mertz2}. Each bright point of the light source deposits an image of the pinhole array on the viewing screen. Knowledge of the 
geometry of the pinholes arrangement (the \emph{coded mask}) allows for an efficient numerical reconstruction of the source~\cite{Calabro}. Initially, random arrays of pinholes, 
used in X-ray astronomy~\cite{Dicke,Horrigan}, were   replaced by binary Uniformly Redundant Arrays (URAs)~\cite{Baumert,Busboom}, which were shown to be optimal for 
imaging~\cite{Fenimore1,Fenimore2,Fenimore3,Fuchs,Lidl,Beth,Bomer,Bomer2,Gottesman1,Gottesman2,Liu}. The peculiar autocorrelated distribution of pinholes allows to contain a 
quasi-uniform amount of all possible spatial frequencies. Thereby, allowing high spatial resolution without limiting the image brightness. Furthermore, more information about the source 
object is encoded in the scaling of the shadow image of the object points, so leading to a stereographic effect. In particular, hard X-ray astronomy commonly uses URA-based coded 
masks~\cite{Gunson,Nugent,nasa} and their generalizations, like MURA matrices (Modified URAs), which will be described in the next Sections, as well as spectroscopy~\cite{Harwit}, 
medical imaging~\cite{LanzaBr,Barrett1,Swindell}, plasma physics~\cite{Fenimore4} and homeland security~\cite{Cunningham}. In the type of applications we are interested in, the light 
sources are posited at length scales  of the optical apparata (from meters down to centimeters), then we are mainly concerned with the so-called \emph{Near Field} settings. They imply 
important geometrical effects, leading to distortions in the collected data and the presence of artifacts in reconstruction of the image. Thus, this situation has to be carefully considered, 
in order to improve the reconstruction technique. 

\subsection{General geometrical settings}
\begin{figure}[h]
     \centering
     \vspace{-0.4cm}
     \includegraphics[width=9cm]{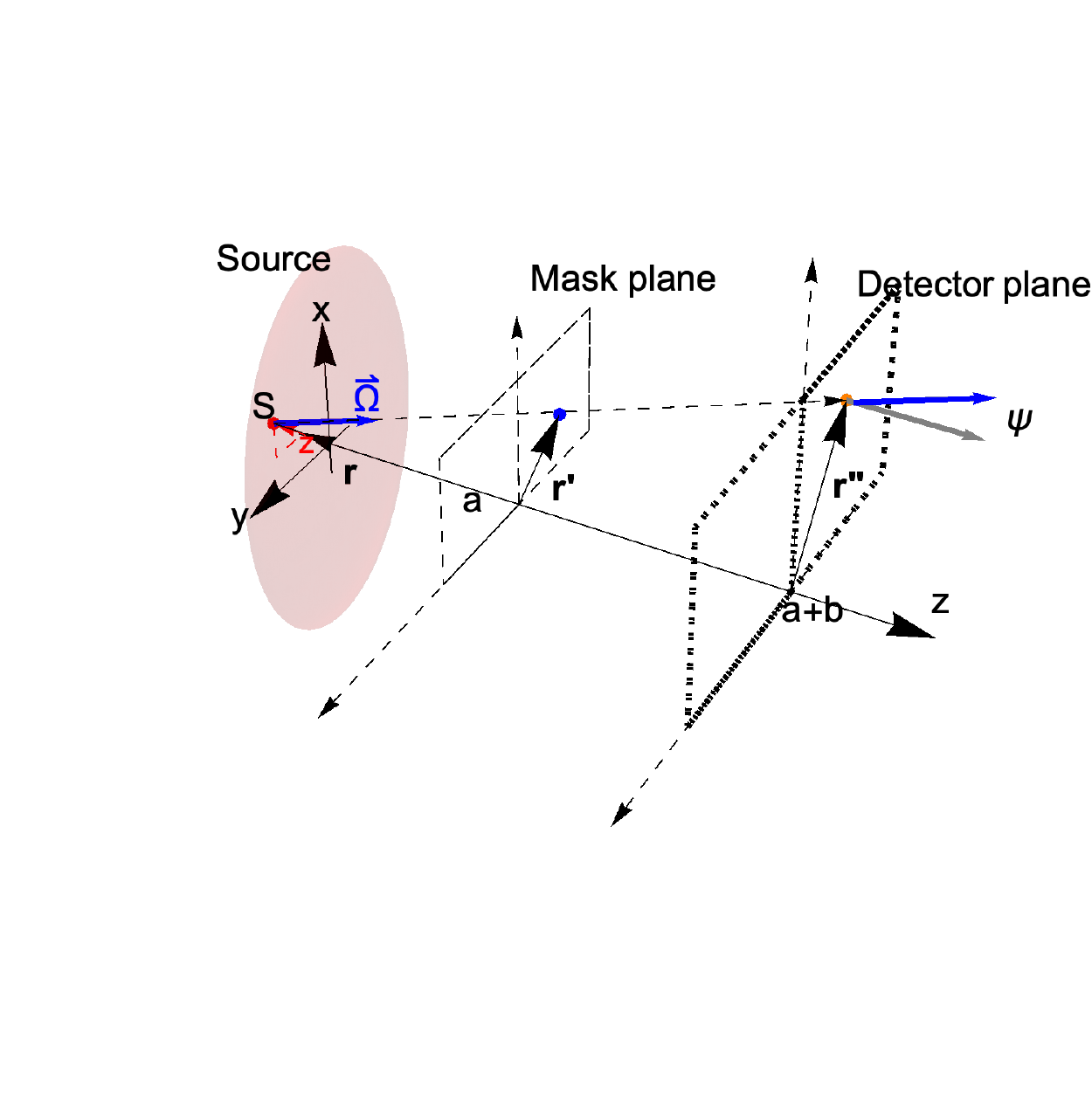}
     \vspace{-2.0cm}
     \caption{Geometry of a ray emitted by the source point $S$ at $(\vec r, z)$ and absorbed by the detector at $\vec r\ ''$ after passing through the coded mask in $\vec r\ '$.} \label{fig:00}
\end{figure}

By scanning the literature, we notice that several simplifying approximations adopted elsewhere, say in astronomy, do not apply in the experimental conditions we assume here. 
More precisely, specific features/requirements of our setup are listed in the following:
\begin {itemize}
\item Near Field sources, that is their typical spatial extension and distance from the detector are of the same order of magnitude as the optical apparatus (typically tens of centimeters),
\item non-planar sources, 
\item filiform sources, 
\item weak sources ($ \approx 10 ^ 4\ photons/sr/cm$)
\item non-static sources,
\item limited detector information capacity ( $10^2 - 10^3$ electronic digital channels),
\item need for a 3-D reconstruction.
\end {itemize}
These settings are mathematically described by a function $\cO\lf \vr ,z, \hO , t\rg$ that denotes the light density of the source to be detected. The variable $t$ is the time and the other ones are 
drawn in Fig.~\ref{fig:00} and discussed in the following. Each point of the source is labeled by the coordinates $(\vec r, z) = (x, y, z)$. The mask and detector planes are parallel to each other and 
are placed along the $z$-axis. A source point $S$, emitting in the direction $\hO$, leaves a projection on the mask plane, whose coordinates are labeled as $\vec r^{\ '} = (x^{\, '},y^{\, '})$, and a 
projection on the detector plane, whose coordinates are labeled as $\vec r^{\, ''} = (x^{\, ''}, y^{\, ''})$.

Moreover, we assume that  the diffraction effects can be neglected, as the aperture size $p_m$ of the single mask pixel is sufficiently large ($p_m \gg \lambda_{VUV}$) to make  the geometrical 
optics approximation still good. Also, interference effects of light coming from the different apertures are neglected. These strong assumptions will be verified in future works, distinguishing them from genuine
noise effects. Further, we assume light to be monochromatic, disregarding at the first stage the effects of finite band width in the spectrum of the emitted light.

In an approximate modeling of the imaging  phenomenon  in a perfectly transparent medium (for instance see~\cite{Accorsi2}), we further assume a planar, isotropic and time-independent averaged 
density of the emitted photons at $\lf \vr , z \rg$, thus
\begin{equation}
\cO\lf \vr ,z, \hO , t\rg= \frac{O_0\lf \vr \rg}{ 4 \pi}  \delta\lf z \rg.
\end{equation}
This source provides an image on a plane detector placed at the distance $a+b$ from the reference frame origin, where $a$ is the focal plane-mask distance and $b$ is the detector-mask 
distance. At the point $\vr^{\,''}$  on the  detector plane $z = a +b$, the image is described by the collected density of photons $P\lf \vr^{\ ''} \rg$ and it is provided by the  integral linear mapping 
\beq 
P\lf \vr^{\,''} \rg \propto  \int_{Source} O \lf \vec{ \xi } \rg \, A \lf \frac{a}{a+b}\lf \vr^{\, ''} + \vec{ \xi}\rg \rg \, \lq 1+\frac{ |\vr^{\, ''} - \frac{a}{b}\,  \vec{ \xi\ }|^2}{\lf a+b\rg^2} \rq^{-\frac{3}{2}}  d  \vec{ \xi}\, , 
\quad
\label{genfield}
\eeq 
\begin{equation*}
\vec{\xi}  =  \frac{b}{a}\ \vr ,
\end{equation*}
where the scaled variable  source density  $O\lf   \vec{\xi } \rg = O_0\lf \frac{a}{b} \vec \xi \rg$ is filtered  by a  kernel, which is the product of a geometrical projective factor and the transmission 
\textit{aperture mask function} $A \lf \vr^{\,'} \rg  , $ where $\vr^{\,'} = \frac{a}{a+b}\lf \vr^{\, ''} + \vec{ \xi}\rg$ denotes the points belonging to the mask plane at $z = a$. Typically, $A \lf \vr^{\,'} \rg$
is a function taking values on $\lgr 0,1 \rgr$ on the mask plane (supposed to be parallel to the detector plane), whose domain is the union of non intersecting squares of equal side length 
$p_m$, defining the apertures of the mask. The values $1$ correspond to apertures and the values $0$ to blind regions.

Ultimately, the function $A \lf \vr^{\,'} \rg   $ is completely defined by a binary matrix, denoted by $a(i, j)$, of suitable dimensions $q_x \times q_y$ (not necessarily equal to each other) 
corresponding to the optically useful region. Thus, a point-like source located in $\vr$ on the $z= 0$ plane contributes to the image, if the vector $\vr^{\,'}$ is such that $A \lf \vr^{\, '} \rg = 1$. 
Moreover, if  a whole mask aperture is illuminated by  a point source, the projected image on the sensor screen will have the size $p_m (a + b)/a$.

The expression~\eqref{genfield} further simplifies in the \emph{Far Field } approximation, {\emph{i. e.}} $|\vr^{\ ''} -  \vr |  \ll a+b$, and accordingly the geometrical projective factor reduces to 1. 
Thus, one obtains the image function in the Far Field form as
\beq
P^0\lf \vr^{\, ''} \rg = \lf O \otimes A \rg \lf \vr^{\, ''}\rg =  \int_{Source} O\lf \vec{\xi} \rg  \, A\lf \frac{a}{a+b} \lf \vr^{\, ''} + \vec{ \xi} \rg \rg  \, d  \vec{ \xi}.
\label{farfieldred}
\eeq
\noindent When $\frac{a}{b}\frac{| \vec{ \xi}|}{|\vr^{\, ''}  |} = \frac{| \vr |}{|\vr^{\, ''}  |} \lesssim  0.1$ 
the paraxial approximation still holds and, in {\color{red}~\eqref{genfield}},  one can resort to a truncated Taylor series expansion of the  geometrical projective factor around $  \frac{| \vr |}{|\vr^{\, ''}  |} = 0$. In most of the 
applications, the expansion up to second-order~\cite{Accorsi1} is considered, but we will limit ourselves to the zero-order, thus providing a $\vec{ \xi}$-independent  distortion of the image. 
Thus, due to the finite source/detector distance,   the geometrically distorted density can be approximated by a  corrected  correlation~\eqref{farfieldred} according to
\beq P^{corr} \lf \vr^{\, ''} \rg =  {\left(1 + \frac{|\vr^{\, ''} |^2}{(a+b)^2}\right)^{3/2}}\; P^0\lf \vr^{\, ''} \rg,\label{distorted} \eeq
which can be used in the reconstruction with the same procedure. In the setting we are going to consider, the correction introduced 
by the prefactor in (\ref{distorted}) is a function of $\vr^{\, ''}$, radially increasing its value up to 5-6\% at the border of the mask with respect to the value at its centre. This geometric distortion 
of the collected intensity of the image may be of some relevance in the case of long tracks, crossing the field of view. Otherwise, for paraxial sources of angular  apertures $< 10^{\circ}$, the 
correction may be completely discarded.
  
\subsection{Focal plane}
The quality of the imaging process is critically determined by the technological characteristics of the photodetectors, intended for the capture and recording of photons 
arriving at the detector. Without going into further details, let us assume that the sensitive region is covered by a square grid of pixels, each with a $p_d$ side.
  
A crucial aspect of the coded mask imaging is the existence of a special plane, called \textit {focal plane}, parallel to both the mask and the detector planes. It emerges 
by observing that, in general, the projection of a mask aperture does not cover exactly an integer number of pixels. In fact, the size of the aperture shadow depends on the 
factor $(a+b)/a$ and it is projected on a number 
\beq
\alpha = \frac{a+b}{a}\ \frac{p_m}{p_d}
\eeq
of detector pixels. Because of the discrete character of the coding  procedure,  in order to avoid generic fractional covering of the photosensors, which  will lead to  defocusing and  
artificial  effects in the reconstruction, it is clear that  $\alpha$ has to take  only integer values. Thus,  a focal plane corresponds  to take $\alpha = 1, \; 2, \; \dots$  and, correspondingly, 
it determines  the distance source-mask $a$, if all the other parameters are fixed by technological requirements. On the other hand, $a$ is  constrained  by the physics we are interested in. 
Typically, we will privilege the plane corresponding to $\alpha = 1$.
  
The aspect to outline in this context is that the process of reconstruction provides a  representation of the light source on the focal plane. Thus, for our research, the particle 
filamentary tracks are directly reproduced only when they lay on the focal plane, or cross it at a small angle.  Then, a question to be answered is how to estimate the focal 
depth of the coded system and which are the corrections to be implemented, to obtain a  suitable reconstruction.
 
A further consequence of this geometrical setting is the concept of \textit{Field of View} (FoV), defined as the portion of the focal plane that projects the entire pattern of the mask on a 
finite size detector. Equivalently,  one may consider the counterimage on the focal plane seen by a single aperture. That will be a square of side length $l_m =  p_m (a+b) / b$. Hence 
the FoV is simply a rectangle of area $ (q_x \times l_m) \times (q_y \times l_m)$, where $q_x$ and $q_y$ are the number of rows and columns in the coded mask matrix, respectively. 
In this perspective, the focal plane is a covering of a set of squares (cells) of minimal side length $l_{res} = p_d\ a /b$, each of them projected one-to-one on a detector pixel of area 
$p_d \times p_d$. Thus, $l_{res}$ is  the \textit{resolution length} of the system, whose evaluation for $b/a \ll 1$ is $ l_{res} \simeq l_m \label{minLBo}$. The parameter  $l_{res}$ is 
particularly relevant, since it predicts the ability  to distinguish different sources in the FoV. Furthermore, $\theta_{FoV}$ is the angle under which $l_m$ is seen by the detector plane.

\section{Decoding: general aspects}
Our aim is  to decode the experimental image $P\lf \vr^{\, ''}\rg$, also in the corrected form~\eqref{distorted}, in order to reconstruct the source function 
$O\lf \vr\rg$. To this purpose, if formula~\eqref{farfieldred} still holds, we need to find a suitable kernel function for the decoding operator $G$, such that
$ ( A  \otimes G ) \lf  \vr \rg = \delta \lf \vr \rg$.  Thus, the reconstruction problem of the source function in terms of the given image is  ruled by
\bea     
& \lf P \otimes G \rg \lf  \vec{\xi} \rg  = 
(O   \otimes A )\lf  \vr{\,}'' \rg \otimes 
G\lf  \vec{\xi}\rg \propto \nn \\  & O\lf \vec{\eta}  \rg \star \lf   
A  \otimes G\rg \lf \vec{\xi} - \vec{\eta}\rg =  O\lf \vec{\eta}\rg
 \star \delta  \lf\vec{\xi} - \vec\eta\rg =  O \lf \vec{\xi} \rg, 
\label{decodingFormula2}
\eea
where $\star$ is  convolution product.

As seen above, the mask function $A\lf \vr\ ' \rg$ naturally introduces a discretization, described by the matrix $a\lf i, j \rg$  and the scale parameter $p_m$. Therefore one has to 
look for a discrete version $g\lf i, j \rg$ of the decoding kernel $G$, accompanied with suitable correlation conditions of the form 
\bea
\sum_{i=0}^{q_x -1}\sum_{j=0}^{q_y -1} a\lf i, j \rg g\lf i +l , j +k \rg  \approx  \delta\lf l, k \rg.
\label{eq:42}
\eea 
 
To this aim, it was shown in~\cite{Fenimore1,Fenimore3} that an optimal compromise between the reduction of \emph{coding noise} (or \emph{artifacts}) 
and the amplification of coherent effects, known also as \emph{discretization noise}, is obtained if  the \emph{periodic autocorrelation function}  (PACF) of 
the aperture array has constant sidelobes, {\emph{i. e.}}
\beq \phi\lf l, k\rg = \sum_{i=0}^{q_x -1}\sum_{j=0}^{q_y -1}
 a\lf i, j \rg a\lf i +l \; \textrm{mod}\, q_x, j +k \; \textrm{mod}\, q_y \rg= \lgr 
\begin{array}{cc}
  K&   \lf l, k\rg = \lf 0, 0\rg   \\
  \lambda&   \textrm{otherwise}
\end{array}
\right. , \label{sidelobes}\eeq
where the peak $K$ and the sidelobe parameter $\lambda$ are numbers to be determined.  Therefore, by combining~\eqref{sidelobes} and~\eqref{eq:42} it is very easy to
compute the decoding kernel of $G$, which will be of the form $g\lf i, j \rg = \frac{a\lf i, j \rg}{K-1} - \frac{\lambda }{K \lf K-1\rg}$. Arrays with this property are commonly referred 
to as \emph{Uniformly Redundant Arrays} (URAs), as  originally introduced by Fenimore and Cannon~\cite{Fenimore1} for the special case $q_y = q_x +2$ both prime integers.
The  construction of such a family of matrices  is based upon quadratic residues in \emph{Galois fields} $GF(p_1, \dots , p_n)$ ($p_i$ are integer powers of prime integers)~\cite{Busboom}.
We consider here a slight variation of URAs, called \emph{Modified Uniformly Redundant Arrays} (MURAs)~\cite{Gottesman2,Busboom}, which  is a family of  arrays obtained by
the method of the quadratic residues for $q_y = q_x = q$ prime integer, but possessing a PACF with two-valued sidelobes, {\emph{i. e.}} $\lambda_1$ and $\lambda_2$, instead of 
a single one\footnote{$\lambda_1 - \lambda_2 =1$}. For large $q$ it can be proved that the ratio (called the \emph{open fraction}) of the apertures with respect to the total number 
of the matrix elements rapidly tends to $50\%$. 

\begin{figure}[h]
     \centering
     \includegraphics[width=8cm]{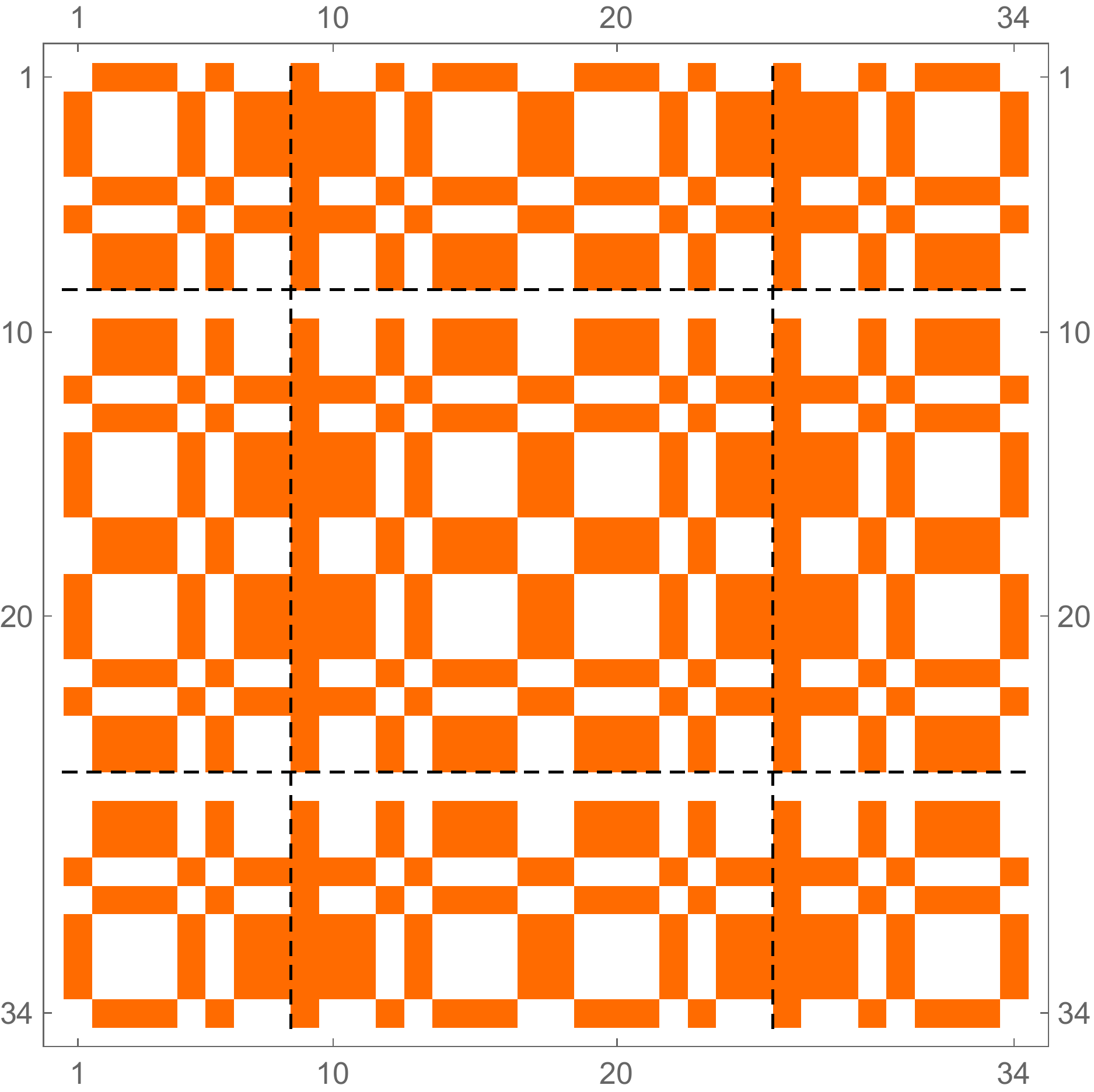}
    % \vspace{-1cm}
     \caption{A mosaic of four  MURA $17\times 17$ matrices, after global  permutations both of raws and columms. Black dashed lines delimit a basic central mask, the other three are  spread in the remaining sectors.    The orange pads correspond to entry 1 and the white ones correspond to 0.} \label{fig:17x17}
\end{figure}

MURAs offer the advantage to be square matrices with open fraction  $\sim 50\%$, furthermore the algorithmic   construction and the decoding kernel of $G$ are simple modifications 
of the URA's case. Since the construction method  of the MURA masks  is well known from the literature, here we report only the basic formulas for  a   $q \times q$ matrix, 
\begin{equation}
  a\lf i, j\rg  = \lgr 
  \begin{array}{ccc}
 0 & i = 0 &    \\
 1 & i\neq 0  &  j = 0   \\
1  &   i, j \neq 0 & a_1\lf i\rg = a_1\lf j\rg\\
0   & i, j \neq 0 & a_1\lf i\rg \neq a_1\lf j\rg
\end{array}, \quad g\lf l, k \rg = \left\{
\begin{array}{cc}
1 & l = k =0  \\
2  a\lf l, k\rg -1 & l+k\neq 0 
  \end{array}
\right.
\right. , \label{MURA} \end{equation}
 where $a_1$ is a Legendre sequence of order $q$, given by 
 \begin{equation}
  a_1(i) = \lgr 
  \begin{array}{rl}
  0 &  i = 0  \\
 +1 &    i = \textrm{mod}_q \mu^{2j} \\
 -1 &    i = \textrm{mod}_q \mu^{2j+1}
\end{array}
\right. ,\label{legendre} \end{equation}
for a generating element $\mu$ of $GF(q)$ (see Fig.~\ref{fig:17x17}). To enlarge the FoV,  we will consider combinations  (mosaic) of masks, assembled side by side in juxtaposition and possibly with rows and columns cyclically permuted, which do not change the PACF function. 

Once obtained  the image of a source on the detector screen $P_{det}$, it will be decodified by using a suitable discretization of the formula 
in~\eqref{decodingFormula2} and the deconvolution  matrix in~\eqref{MURA}, which we reproduce here for convenience
\beq 
O_{l , k} = \sum_{i=0}^{q -1}\sum_{j=0}^{q -1} P_{det} \lf i,  j \rg \;  g\lf{(i + l)\hspace{-0.3cm}_{\mod_q} +1}, {(j + k)\hspace{-0.3cm}_{\mod_q} +1 }\rg \label{deconvolution}.
\eeq

\noindent Using such a general procedure, one may manage simulations, up to now only geometrical but meaningful (Sec.~\ref{sec:simu}), of several light signals 
for testing the general properties of the coded mask technique. 

\subsection{Integer Affine Transformations}
In order to extract the basic properties  of the imaging process via coding masks,  we introduce here an algebraic approach, based on the observation that the image of a point-like source $S$ 
on the detector can be represented as an inhomogeneous affine mapping, from the mask points  to the  detector points, parametrically dependent on  the $S$ coordinates. After a suitable change
of the reference frame, the source is coordinated by $S = \lf \mu_S\, p_m , \nu_S\, p_m  , \delta_S\, a\rg$, in terms of the aperture pitch $p_m$ on the mask plane and the focal distance $a$. 
Specifically, along the axis orthogonal to the mask, the third coordinate $ z = \delta_S\, a$ is expressed by the relative distance $\delta_S$ of the source from the focal plane. Thus,  the source is located between 
this plane and the mask for $ 0 <\delta_S <1 $, otherwise it lies beyond it  for $ \delta_S> 1 $. In its turn,   $S_\bot= p_m \lf \mu_S, \nu_S \rg$ expresses the orthogonal projection of the source 
position on the mask plane. Likewise, the mask apertures will be denoted by   the set of  coordinates $H = p_m \mathcal{H} $,  where $\mathcal{H}$ is a list of ordered pairs of integers 
$\lf \mu , \nu \rg$ only,  identifying  a  lattice of point-like apertures on the mask. Thus, assuming the point-like source $S$ on the focal plane,  we can represent the discretized 
image density as  the affine transformation over a  $q^2$-dimensional vector space (from the mask lattice) by 
\bea T_{S} {H}  = - \frac{b}{a} S_\bot + \lf 1 + \frac{b}{a} \rg H   =  p_m \lfq - \frac{b}{a} \lf \mu_S, \nu_S \rg +  \lf 1+ \frac{b}{a} \rg \mathcal{H} \rgq,\label{mapping1}\eea
which singles out a lattice of points on the detector plane. Since the blind pixels in the mask are excluded from the mapping, actually   only  a  number, equal to the peak value of the mask $K = \frac{q^2-1}{2}$, of correspondences is needed to be computed. 

Since the above mapping is equivalent to projecting only one light ray from the source to the detector through a single point (for instance its center) of the aperture,  it is more realistic to consider more of such 
points, in particular,   closer to the aperture sides. Then, one may consider  $\rho + 1$ crossing points for each aperture, generating the set of coordinates 
\beq 
\left.
\begin{array}{l}
  \ \mathcal{H}_{\sigma} = \lgr   \lf \mu +\sigma_1^i, \nu + \sigma_2^i \rg  , \quad 
  \, |\sigma_k^i| \leq \half  ,\quad i= 0, \dots \rho  ,\quad k = 1,2 \rgr.
\end{array}
\right. \eeq
 Moreover, as mentioned above, in order to widen the Field of View (FoV),  a mosaic of four masks can be  implemented, further enlarging the set  $\mathcal{H}_{\sigma}$. Hence, the previous 
 mapping~\eqref{mapping1} can be modified for sources in  generic positions and rescaled by the pitch $p_d$ of the detector pixel  as follows 
\beq 
\widehat{T_{S}}  {H} = { \frac{\; p_m}{ \;p_d} \lfq - \frac{b}{\delta_S\ a} \lf \mu_S, \nu_S \rg + \lf 1 + \frac{b}{\delta_S\ a} \rg   \mathcal{H}_\sigma \rgq}.  \label{mapexact}
\eeq
Finally, since the detector pixels are also quantized and identified by a lattice of pairs of integers (in $p_d$ units), each projected point $\widehat{T_{S}  }  {H} $ is properly  assigned  to a specific  pixel by rounding up
\beq
\widehat{T_{S}} {H} _{device} =  \lfq\widehat{T_{S}} {H}  \rgq,\label{mapdev}
\eeq
which  is  a nonlinear and not invertible operation. Thus, part of the complete information will be lost and an intrinsic discretization noise is introduced. So, a statistical analysis may be required 
to deal with the detected data. Thus, the representation \eqref{mapexact}-\eqref{mapdev} of the coded mask action allows us to algebraically study some of the main properties of  the acquisition and 
reconstruction algorithms.  In particular, one can  obtain a dual spectral description of the masks. To this aim, let us observe that: 1.  the 
individually resolved sources  belong to the lattice of points on the FoV $\lgr S_{\bot\, i, j  } \rgr=  l_{res}  \times \lgr \lf i, j \rg \rgr$ , with  $i, j$ running over  $1 - q$,  and  2. the relation \eqref{mapexact} is linear in $S_\bot$. Thus the $q^2 \times q^2$ matrix  of columns $\Phi = \lgr \widehat{T_{{S_1} }} {H}, \dots , \widehat{T_{{S_{q^2}} } }{H} \rgr$ represents a linear application  from the $q^2$-dimensional  space of the discretized light distribution, where  $\lgr S_{\bot\,k  } \rgr_{k = 1, \dots , q^2}$  forms a basis in that space. The  target  discretized  image space $Y$ represents the pixel measurements on the discretized device plane.  Thus, one handles with a fully discretized version of the coded mask
transfer matrix for the linear mapping 
\beq Y = \Phi\; S_\bot + E \eeq
 possibly encoding the intrinsic and extrinsic 
noise into the vector $E$.  Multiple point-like sources  superimpose their images which,  because of the limited resolution power or by
  the discretization,   are defocused on two or more surrounding  device pixels. Moreover, here it is important to notice from \eqref{mapexact} that the dependency of  $\Phi$ on the  relative distance $\delta_S$ is non linear and needs a separate discussion. Actually, in the on-focal-plane case the relations~\eqref{mapexact}-\eqref{mapdev}  provide the transfer matrix $\Phi$, which could be algebraically derived from its very definition in terms of the MURA mask.  It turns out that   $\Phi$ is a binary symmetric non degenerate matrix, its inverse  represents the action of 
the corresponding decoding operator $G$, and its  spectrum is real and by induction can be proved to be
\beq 
\sigma\lf \Phi \rg = \lgr \frac{q^2 -1}{2} = K \, \lf deg = 1 \rg, \pm \frac{q+1}{2} \, \lf deg = K/2  \rg , \pm \frac{q-1}{2} \, \lf deg = K/2  \rg\rgr, \label{spectrumFocal}
\eeq   

\noindent where the eigenvalue degeneracy is given. As a consequence, any combination of  sources in the focal plane is decomposed in the sum of  $q^2$  eigenvectors
\begin{equation*}
\lgr \hat{e}_{\lambda}^i, \quad \lambda \in \sigma\lf \Phi \rg, \quad i = 1, \dots deg\lf \lambda \rg \rgr
\end{equation*}
of $\Phi$. Their images are simply scaled, or reflected-scaled, only by the two distinct, but very close,  factors $ \pm \frac{q\pm 1}{2}$ (a quasi-flat spectrum is a remarkable property allowing for good reconstructions), except for the non-degenerate 
eigenvalue $K$. The corresponding eigenvector, say $\hat{e}_0$, has all equal components, without specific information about  the details of a generic source. However, 
since the other eigenvectors contain negative components, implying "unphysical" sources,   $\hat{e}_0$ is needed to correctly reconstruct the source.
 
On the contrary, out of the focal plane ($\delta_S \neq 1$), numerics is needed to deal with the prescribed relations~\eqref{mapexact}-\eqref{mapdev}.  By using the same lattice of single 
sources as above, the matrix $\Phi$ loses the previous simple structure: it is not symmetric anymore and its elements take values over a finite set of real positive numbers. However, at least 
for the  explored values $\delta_S \approx 1$, these matrices are still diagonalizable, but their eigenvalues take complex values and are not degenerate. Still, there exists a maximal isolated 
real eigenvalue, the others appear in conjugated pairs, whose absolute values fill a band, extending from the degenerated values indicated in~\eqref{spectrumFocal} to 0. So, the spectrum 
is not longer quasi-flat and the phases make the eigenvalues migrate in the disk  around the origin of the complex plane  of radius $\approx q/2$. This corresponds to  a superposition of many 
scalings and rotations of the image around the axes of the optical system.  Even if at the moment we do not have any analytic tool to describe such a situation, remarkably a $\delta_S$-dependent 
rescaling of the source lattice allows to find a pure real spectrum for $\Phi$, which becomes symmetric, but still the eigenvalues range over a band of the order $q$ near 0. The rescaling is of 
the order of $\delta_S$, even if its exact expression for restoring the ideal  simple spectrum \eqref{spectrumFocal} is not achieved yet. Several different techniques to calibrate such a factor 
are actively under investigation with the aim to realize a numerical focusing method.

\section{Design specifications \label{DSpec}}
Since the photosensors will be arranged in a square matrix, we will consider MURA coded masks. Even if conceptually  this is not necessary, the below defined geometries are a compromise among
the technological limitations (available matrices of SiPM photosensors, electronic and mechanical constraints, allowed heat dissipation rate in the scintillation liquid)  and the image reconstruction
requirements. In other terms, we would explore here imaging systems involving few sensor channels, which in a more generic context may be an arbitrarily scalable factor. As mentioned above, to 
enlarge the FoV,  we consider a mosaic of masks. 
 After inspecting several types of assembling, we arrived at the conclusion that the best solution consists of four 
cyclically arranged masks. By exploiting the cyclic shift property of the MURAs, we periodically permute columns and rows also on the mosaic to optimize the resolution of the paraxial light sources. 
Simply, the so built mosaic allows us to expand the region of light collection with the same basic pattern. Thus, sources at large angular position with respect to the normal at the mask can project on 
the detector screen their images coming from different apertures. We have to stress that the detector array keeps the same number of rows and columns as a single mask. Furthermore, after a restriction 
of the image on the effective detector matrix, the deconvolution procedure will proceed as usual. 

\subsection{The 6 detectors setup \label{stage}}

\begin{figure}[h]
     \centering
     \includegraphics[width=8cm, height=8cm]{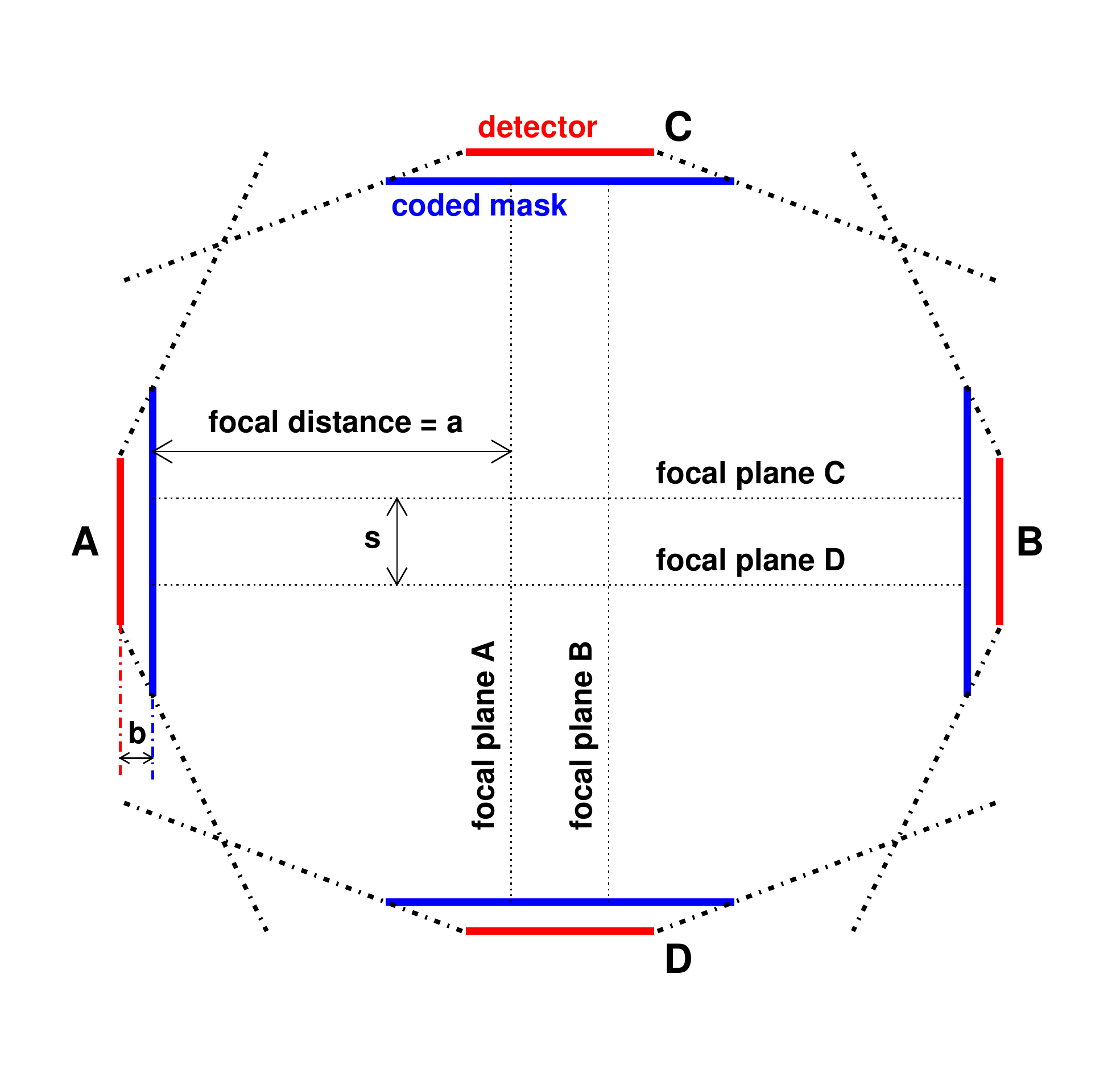}
     \caption{Possible setup with couples of parallel devices (mask-mosaics and detectors). The distances $a$ and $b$ are for mask-focal plane and
                   mask - detector, respectively. In this work the setup with $s=0$ (coincident focal planes) is used for the simulation. 
                   Also a setup with $s<0$ could be designed.} \label{fiducial}
\end{figure}

Most of the previous considerations on the use of  the coded masks of small rank  and in the near field conditions suggest that their stereographic properties are partially shadowed
in the reconstruction of a source. Thus, the obvious solution is to expand the  detector dimensions or, alternatively, try to dispose more of them in different configurations, allowing to
detect the true spatial extension of the  tracks we are looking for. Thus, we arrive at the concept of  a spatially distributed system of coded masks. In particular,  in the present paper we 
propose to consider a Stage of Observation, bounded by pairs of coaxial parallel coded mask devices, as schematically represented in Fig.~\ref{fiducial}. The main
features of such a setup are:
\begin{enumerate} 
\item 6 mask mosaics define a cubic Stage for the physics of interest;
\item the masks are identical in a $2\times 2$ mosaic;
\item each pair of parallel mosaics shares the same symmetry axis;
\item the 6 SiPM detectors are coplanar to the mosaics, at the same distance $b$ from the coupled mask;
\item the center of the Stage is the origin of an orthogonal  reference system;
\item the masks are as far apart from the origin exactly as the focal distance $a$, thus the coordinate planes passing through the origin are themselves  focal planes.
\end{enumerate}
Furthermore, one can outline several details, namely:
\begin{enumerate}
\item the detectors will provide redundant information, which has to be simplified/exploited;
\item a lower number of devices can be used, exploiting more efficiently the performed measurements;
\item a primary interest will be to study the possible measurements by means of couples of parallel coded masks;
\item a second step is to exploit the performances of couples of orthogonal masks;
\item since presently the geometry of the specific experimental imaging-system cannot be fully determined, the cubic setup can be deformed in a more general 
parallelepiped structure, with non-coincident focal planes and shifted masks. 
\end{enumerate}

\noindent Following the previous prescriptions, many different experimental setups have been studied, taking into account actual technical requirements, related 
to the intensity of light emission, number of electronic channels per detector array and geometry of the Stage of Observation. However the analyses presented 
in this paper are based on the simulation performed according to just one design. The parameters of a single device are reported in Table~\ref{tab:prima}.

\begin{table}[b]
\caption{Features of a single device (mask-mosaic and detector).} \label{tab:prima}
\centering
\begin{tabular}{lc} \hline\noalign{\smallskip}
MURA mask                                                                   & $17\times 17$ \\
mosaic of masks                                                            & $2\times 2$ \\
mask aperture side length ($p_m$)                               & $3.15\ mm$ \\
mask - focal plane distance ($a$)                                  & $250\ mm$ \\
mask - detector distance ($b$)                                       &  $20.0\ mm$ \\
detector (SiPM) matrix						   & $17\times 17$ \\
acquisition channels                                                       & $289$ \\
SiPM pixel pitch ($p_d$)                                                & $3.4\ mm$ \\
focal plane separation ($s$)                                          &  $0\ mm$ \\
magnification on the focal plane ($\alpha$)                   & 1 \\
detection efficiency                                                        & $100\%$ \\ \hline
resolution length ($l_{res}$)                                           &  $\sim 13.5\ p_m = 42.5\ mm$ \\
Field of View (FoV)                                                                  &  $\sim 723 \times 723\ mm^2$ \\
angular aperture ($\theta_{FoV}$)                                 &  $\sim 8.9^\circ$ \\ \hline
\end{tabular}
\end{table}

\section{The Single Pinhole Camera Approximation \label{harmonicmean}}
The problem of the spatial localization of the source is hardly solved by using only one coded  small-order mask, then we need to use more than one. The 
simplest considered mask arrangement is made up by two parallel coded  systems, sharing the same focal plane. For sake of simplicity, we suppose that the 
apertures of the two masks result to be co-axial, that is, any orthogonal line to mask planes intersects the corresponding apertures on both of them. However 
the results we are going to present can be extended also to non-aligned masks.
 
\begin{figure}[htbp]
\begin{center}
\hspace{-1.4cm}
\includegraphics[width=7cm]{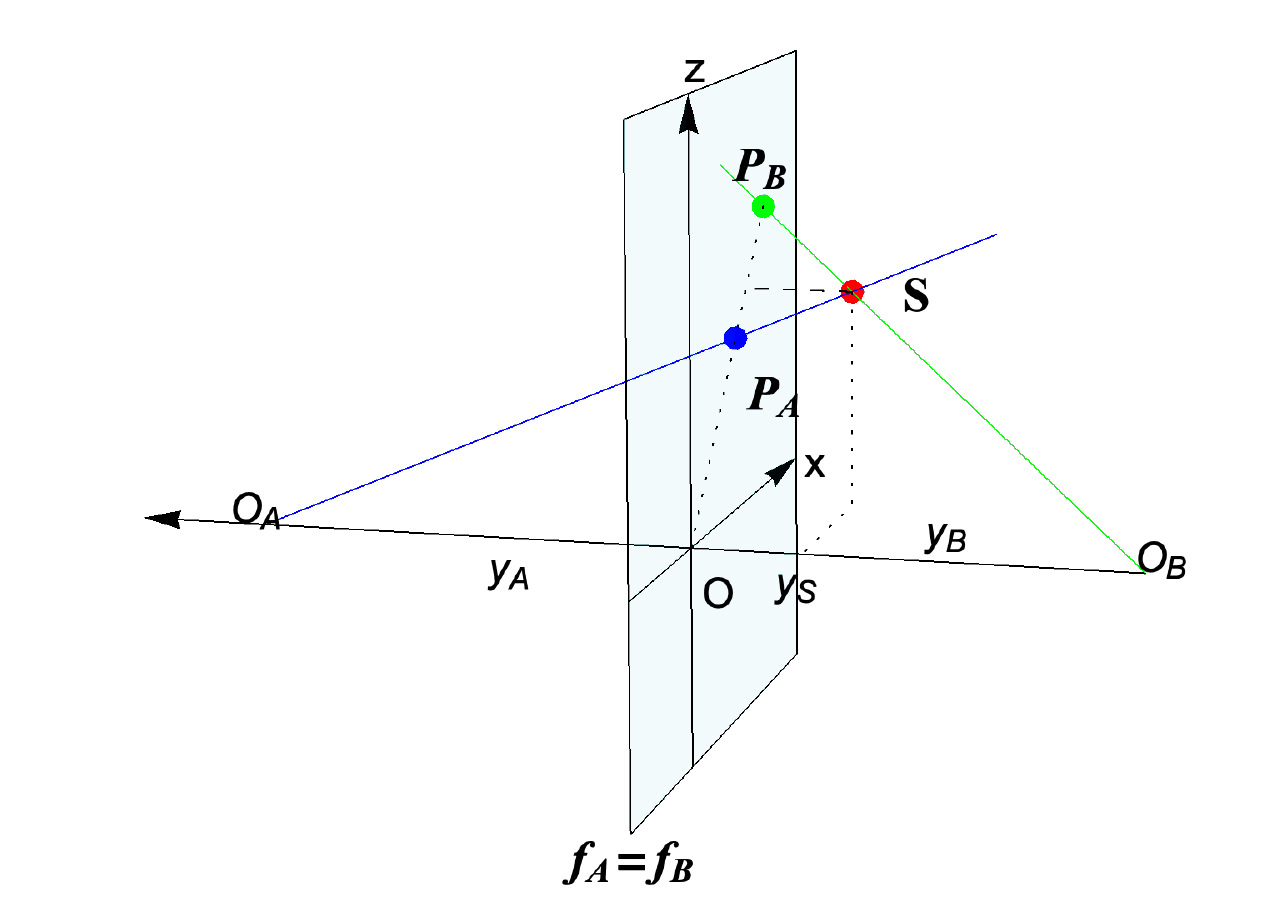}
\includegraphics[width=6.4cm]{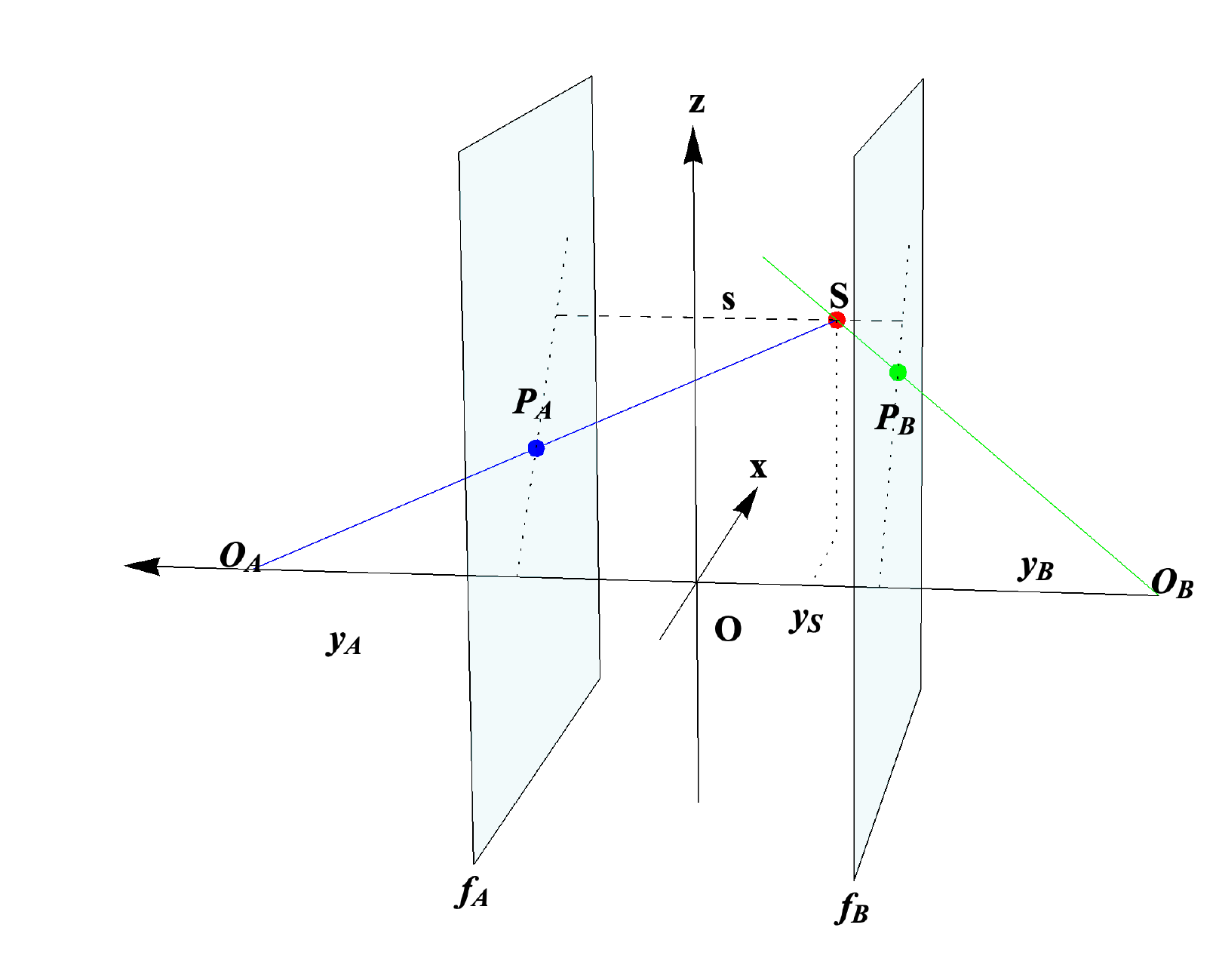}
\caption{\emph{Left} - A source S seen by a pair of  single-pinhole cameras sharing the focal plane ($f_A = f_B$). $P_A$ and $P_B$ are the apparent positions 
observed by $O_A$ and $O_B$, respectively. \emph{Right} - A source S seen by a pair of  single-pinhole cameras with distinct focal planes ($f_A \neq f_B$), 
separated by a supplementary distance $s$. Again, $P_A$ and $P_B$ are the apparent positions observed by $O_A$ and $O_B$, respectively.} \label{dPinHCam}
\end{center}
\end{figure}

In order to obtain simple formulas to reconstruct the image, we approximate each coded mask with a single pinhole camera. Such an effective (point-like) 
aperture is set at the center of each mask, in the origin of the reference frame of the mask (see Fig.~\ref{dPinHCam}, left). Let us call $O_A$ and $O_B$ 
the center points of the masks on the left (A) and on the right (B), respectively,  with the coinciding axes.  In this scheme the imaging process is reduced to 
a projective application of the source points on the focal plane  through the poles  $O_A$ and $O_B$. For the stipulated approximation to be valid, the source 
must be sufficiently far from the mask and the angle subtended by two different apertures of the mask has to be small. Then, taking into account the existence 
of a preferred focal plane, the approximation validity interval can be expressed as
\beq
L \ll  \frac{a^ 2}{q \ p_m},
\eeq
where $q$ as above is the dimension 
of the mask and $L$ is a typical transverse distance of the source from the center of the FoV.
  
In practice, let us fix   a point-like source $S$ located in the space between the two masks, with coordinates $\lf x_S, y_S, z_S\rg$ with respect to a right-handed 
frame of reference $\lf O, x, y, z\rg$. We denote by $y_A = |y_S- a|$ and  $y_B = |y_S+a|$ the length of the projections on the $y$-axis of the segments $\overline{O_A S}$ 
and  $\overline{O_B S}$ with  the restriction 
\beq y_{{A}}+y_{{B}} = 2 a.\eeq
The projection of $S$ on the focal plane is done by the intersection of two straight lines of  the  bundle through $S$ and $O_{A, B}$, respectively. These intersections
are denoted by $P_A = \lf x_A, a, z_A\rg, \quad  P_B = \lf x_B, a, z_B\rg$, which belong to a line passing through the intersection of the axis system $O$ with the focal
plane. This can be proved by elementary geometry. Of course, the reconstruction of sources closer to a mask is seen more far apart on the focal plane, but closer to the 
center $O$ when seen from the opposite side. The Cartesian equations of these two lines are 
\beq \begin{array}{cc}
  x =  -  \frac{x_A}{a} y + x_A   &  \ \ \ \ \ \ \ \ \ \ z =  -  \frac{z_A}{a} y + z_A , \\ \\
  x =  + \frac{x_B}{a} y + x_B   &  \ \ \ \ \ \ \ \ \ \ z =  + \frac{z_B}{a} y + z_B.       
\end{array} 
\eeq
and their intersections are readly found to be
\begin{equation}
x_S = \frac{2 x_A x_B}{x_A + x_B}, \quad  z_S = \frac{2 z_A z_B}{z_A + z_B}, \quad  y_S = a\frac{ x_A - x_B}{x_A + x_B}=a\frac{ z_A - z_B}{z_A + z_B} . \label{harmonicmean}
\end{equation}
The first two previous equations represent the harmonic mean of the $A$ and $B$ coordinates. Such elementary formulas are of great help in localizing sources. In fact, it is 
enough to compute for the same point-like source the $x, z$ coordinates on the focal plane seen by the two masks and compute their harmonic mean, providing the correct value.

In the more general case of non co-focal masks, in the same approximation (see Fig.~\ref{dPinHCam}, right) analogous formulas hold, namely 

\beq
 x_S = \lf 2 + \frac{s}{a} \rg \frac{ x_A x_B}{x_A + x_B}, 
 \quad 
 z_S = \lf 2 + \frac{s}{a} \rg \frac{ z_A z_B}{z_A + z_B}, 
\eeq
\beq
 y_S = \lf a + \frac{s}{2} \rg  \frac{ x_B - x_A}{x_A + x_B}=\lf a + \frac{s}{2} \rg  \frac{ z_B - z_A}{z_A + z_B}. \label{eq:harmo}
\eeq

\noindent where $s$ is the separation distance of the focal planes. This is a significant design parameter, because the photon collection and 
the spatial resolution depend on it. Also, the configuration with $s<0$ (focal plane closer to the opposite mask) can be implemented.

Several configurations of non planar sources have been simulated and successfully analyzed by calculating the harmonic mean. These 
checks are partially reported in the following (Sec.~\ref{sec:4punti}).

Finally, the method is not particularly useful in the numerical evaluation of the third coordinate ($y_S$ in this case) since its estimate  is affected by large uncertainty. 
This behaviour can be understood (disregarding the effect of the quadratic intensity falling off with the distance) by noticing that the localization procedure of a point-like 
source performed here is equivalent to establishing a one-to-one correspondence between the set of sequences of $2^{q\times q}$ bits and the set of adjacent convex 
3-polytopes, generated by the planes emerging from the sensor devices and tangentially intersecting the mask apertures. The polytopes tile the space in front of the mask, 
but their shape is not regular, nor their sizes. In particular, in correspondence with the focal plane, there is a stratum of polytopes, which are elongated in the orthogonal 
direction about ten times the transversal section size $\sim l_{res}$. Thus, they allow us to determine pretty well the coordinates of the sources in that plane, but very 
roughly in the direction of the mask axes.
 
Assuming that the association of pixels on different parallel detectors is correct, the error on $x_S$ is given by the following formula\footnote{According 
to this result, we observe that this uncertainty is lower in the configuration with $s < 0$ but in this case the single-pinhole approximation can be not suitable.}

\beq
\sigma_{x_S} = \lf 2+ \frac{s}{a} \rg  \frac{\sqrt{x_A^4 + x_B^4}}{(x_A + x_B)^2}\  \tilde\sigma, \label{eq:error}
\eeq

\noindent where $\tilde\sigma =  l_{res}/\sqrt{12}$. Same formula can be used for the second coordinate ($z_S$ in this example).

However, it is necessary to stress that a larger error can be introduced by wrong association of pixels on opposite detectors. This association can be driven 
by topological criteria case-by-case. Here we want to stress that the associated pixels must be on the same Cartesian quadrant and on the same line with 
respect to the system origin (look at the position of $P_A$ and $P_B$ in Fig.~\ref{dPinHCam}, left).

\subsection{3-D reconstruction in the single-pinhole camera approximation} \label{sec:3D}
\begin{figure}[htbp]
\begin{center}
\includegraphics[scale= 0.4]{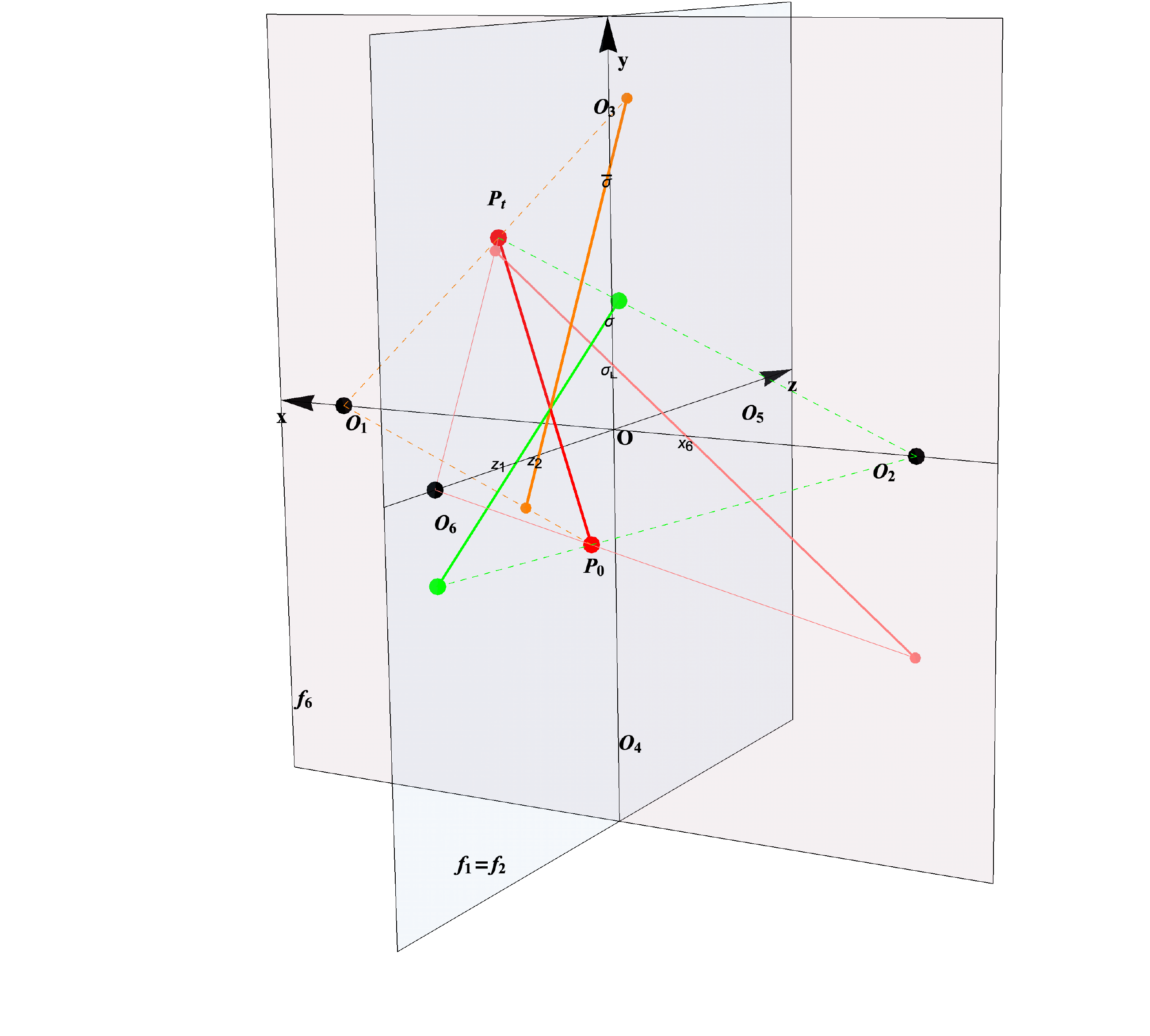}
\vspace{-0.2cm}
\caption{Stereographic projections of a given segment (in red) from three different poles  $O_1, \; O_2, \; O_6$.   $O_1$ and $O_2$ are on the same $x$-axis, 
symmetrically placed  at the distance $\pm a$ from the origin. The third pole $O_6$ is at $-a$ on the $z$-axis. The red segment $\overline{P_0 P_1}$ 
indicates a light track, whose projections are indicated in green and orange on the plane $z-y$ and in  pink on the plane $x-y$. The other visual poles have been 
indicated for future reference.} \label{triprojections}
\end{center}
\end{figure}

By using the single-pinhole approximation, it is quite simple to prove how to reconstruct single linear tracks in the space by means of two parallel co-focal masks
and a third one, this latter placed orthogonally with respect to the other ones (Fig.~\ref{triprojections}). The basic idea rests upon the elementary projective 
geometry.

A physical track, represented by a segment $\overline{P_0 P_1}$, will be stereographically projected by the three line bundles emerging from the pinholes $O_1$, 
$O_2$ onto the common focal plane $y-z$ and from  $O_6$ onto  $y-x$, respectively. The three projected segments belong to straight lines with equations 
\bea 
y &= \bar\rho   \; z + \bar\sigma,    \qquad    &\textrm{on} \;  y-z \qquad  \textrm{from} \; O_1\nn \\
y &= \rho         \; z + \sigma,           \qquad    &\textrm{on} \;  y-z \qquad  \textrm{from} \; O_2  \label{projectline}\\ 
y &= \rho_\bot \; x + \sigma_\bot,   \qquad    &\textrm{on} \;  y-x \qquad  \textrm{from} \; O_6 \nn
\eea
where  $\rho, \;\bar{ \rho}, \; \rho_\bot$ are the angular coefficients and $\sigma,\bar{ \sigma}, \sigma_\bot$ the intersections with the \emph{y}  axis obtained from the 
2-dimensional reconstructed images (see Fig.~\ref{triprojections}). These six quantities are related to the parameters of the physical track in the space, {\emph{i. e.}} its 
direction and one of its points. More precisely, parametrizing  the track line  by the line unit vector  $M = \left(n_x, n_y, n_z \right)$ and one of its point  $P_0 = \lf x_0, y_0, z_0\rg$, 
one can analytically derive the equations of the projected  lines on the chosen planes. Thus, one obtains the parameters in \eqref{projectline} in terms of linear fractional  
combinations of $M$ and $P_0$ components. Thus, it is possible to solve such an algebraic overdetermined  system, together with certain consistency conditions. For 
instance, looking at the first two equations in  \eqref{projectline} and from the observed  projected segments on the $y-z$ plane (detected by $O_1$ and $O_2$), it is easy
to calculate the intercept with the $y$-axis and one of the slope
\bea 
z^* = 0, \ \ \ \ \ \ \  y^* = 2\frac{\bar{\sigma}\ \sigma}{\sigma + \bar{\sigma}}, \ \ \ \ \ \ \  
n_y/n_z = \frac{\bar{\rho} \sigma+\rho  \bar{\sigma}}{\sigma + \bar{\sigma}}. \label{coeffnsup} 
\eea
Thus, the angular coefficient of the track projected on the $y-z$ plane is the mean value of $\rho, \bar\rho$, weighted with the inverted intersections ($\rho$ weighted 
with $\bar\sigma$, $\bar\rho$ with $\sigma$). Remarkably, this result can be also simply deduced taking into account that the projected line intercepts the $y$ axis in the 
point $(z^*, y^*)$ and the $z$ axis in the point with coordinates
$$z^\dag = \frac{-2\ \sigma\ \bar\sigma}{\sigma \bar\rho + \bar\sigma \rho}, \ \ \ \ \ \ \  y^\dag = 0.$$

On the other hand, the information connected with the projected segments on the $y-z$ plane allows to estimate the other component of the line unit  vector along the 
perpendicular axis, accordingly with the analytic formula, namely
\bea 
n_x/n_z = a\ \frac{\bar{\rho} - \rho}{\sigma + \bar{\sigma}}. \label{angcoeff}
\eea
But, as remarked in the previous section for the last equation in~\eqref{harmonicmean}, we expect that the collected data will provide quite inaccurate evaluations of such 
a parameter. In any case, three pairs of parallel masks allow to use equation~\eqref{coeffnsup} cyclically permuting the oriented planes and obtaining:
$$\lgr y-z\ view \rgr \to z^*=0, y^*, n_y/n_z,$$
$$\lgr x-z\ view \rgr \to z^*=0, x^*, n_x/n_z,$$
$$\lgr y-x\ view \rgr \to x^*=0, y^*, n_y/n_x.$$
Of course, one may apply  similar considerations from the  images  projected onto  pairs of orthogonal planes. For the pair $\lgr y-z \rgr_{O_1}, \lgr y-x \rgr_{O_6}$
in~\eqref{projectline}, one obtains 

\bea  
n_y/n_z &=&  \frac{a^2 \bar{\rho } \rho_\bot+\bar{\sigma } \sigma_\bot}{a (\bar{\sigma }+a\rho _\bot)}, \label{coeffnsum}\\
n_x/n_z &=& \frac{a \bar{\rho }-\sigma_\bot}{\bar{\sigma }+a \rho_\bot}. \label{coeffnpsum}
\eea

\noindent Also in this case, we expect that~\eqref{coeffnsum} will be very useful in determining the slope of the line, while~\eqref{coeffnpsum} will  be affected by large 
uncertainty. But, in the perpendicular-mask setup we can invert the role of the detectors in equation~\eqref{coeffnsum} in order to get the slope in the $y - x$ view. So it results

\bea  
n_y/n_x = \frac{a^2 \rho_\bot \bar\rho+\sigma_\bot \bar\sigma}{a (\sigma_\bot+a\bar\rho)}. \label{inverted}
\eea

\noindent Of course, consistency relations arise when comparing various formulas among themselves. Specifically, the following identities hold 
\beq 
a^2 \left( \bar\rho-\rho \right) \rho_\bot + \left(\bar{\sigma}+\sigma \right) \sigma_\bot = a \left( \bar\rho \sigma + \rho \bar\sigma \right),
\qquad 
a \left(\bar\sigma - \sigma \right) \rho_\bot + \left( \bar\sigma + \sigma \right) \sigma_\bot = 2 \sigma \bar\sigma.
\eeq

\noindent These relations can be exploited not only to check the consistency of the data, but also to remove the dependence of the formulas on $\rho$ 
and $\sigma$. So, two perpendicular detectors are good enough to get the intercepts on $y$ axis
\beq 
x = 0 \rightarrow  y = \frac{a \rho_\bot - \sigma_\bot}{a \rho_\bot + \bar\sigma}\ \bar\sigma, \label{p0}
\qquad
z = 0  \rightarrow y = \frac{\bar\sigma - \sigma_\bot}{a \rho_\bot + \bar\sigma}\ a.
\eeq
 
Therefore, a couple of perpendicular detectors are the minimal setup for the 3-D reconstruction of a linear track and we expect that similar algorithms 
can be implemented also for second-degree curves. It is evident that the procedure here presented does not take into account the actual experimental 
obstacles. Then, many detectors in parallel and perpendicular configuration are mandatory to make redundant measurements in large volumes.

\section{Simulation and signal reconstruction}\label{sec:simu}
In order to quantitatively study strength and  weakness of the coded mask system so far described and to get a more realistic 
understanding of the coded masks performance, a toy Monte Carlo has been implemented. The light rays are emitted uniformly along 
the simulated track and propagate linearly according to the direction extracted isotropically on the full solid angle.

At this stage of investigation, we are concerned mainly with the effect of coded masks on the light signal by looking for simple formulas to 
decode it. In order to reach this goal we did not simulate the actual physical effects (light yield from LAr~\cite{yield1, yield2}, light absorption, 
light scattering, SiPM efficiency and so on). In the future a full Monte Carlo simulation will be implemented.

The simulation has been performed with different parameters (mask rank, pixel sizes, values of $a$ and $b$ lengths) in order to 
verify the correctness of the formulas presented in this paper. Here we present the results obtained with the experimental setup made 
by 6 coded masks and 6 SiPM matrices (see Tab.~\ref{tab:prima} and~Fig.~\ref{fiducial}) with coincident focal planes ($s=0$). 
The full optimisation of the detector setup will be defined by also taking into account mechanical and cryogenic requirement (see Sec.~\ref{sec:intro}).

\begin{figure}[hbp]
\begin{center}
\includegraphics[width=12.0cm, height=10.0cm]{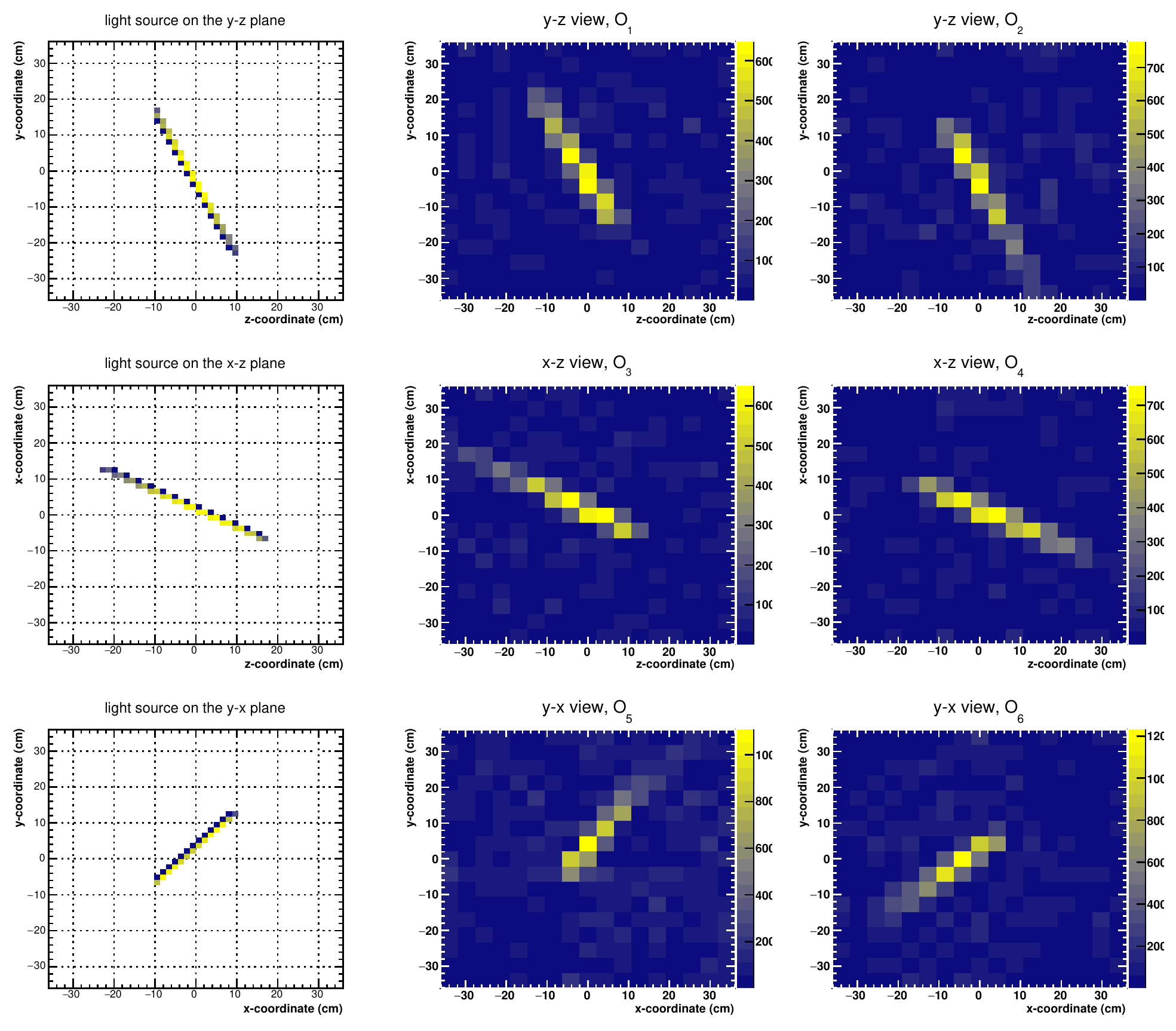}
\caption{\emph{Left column} The light track crossing is shown in real dimensions in three different views. \emph{Central and right columns} Reconstruction 
of the light track on 3 couples of parallel detectors. The experimental setup is that in Fig.~\ref{fiducial}.} 
\label{diagonal}
\end{center}
\end{figure}

\noindent A reference example of the projection/reconstruction procedure applied in the present work is reported in Fig.~\ref{diagonal}, where a simulated 
linear light-track crosses the Stage of Observation. In the left column the original track is plotted in the different views. In central and right columns the track 
is reconstructed using the photon signal collected by the SiPM detectors. At this step we did not apply any kind of filter on the reconstructed image, we simply 
take the absolute value of the content of each bin (as an effect of the decoding also negative values are possible, but significantly lower in absolute value than 
the signal). We would like to stress that the track does not lay on focal planes. Then, as expected, the shape of the reconstructed track in the same view is different 
because of the point-by-point distance of the track from each detector. Indeed the measurement with only one mask reproduces a distorted image of the track. 
A true image can be reconstructed combining the information from different masks, as explained in previous sections and verified in the following ones.

\subsection{Application of the Harmonic-Mean method and signal filter}\label{sec:4punti}
\begin{figure}[htbp]
\begin{center}
\includegraphics[width=12.0cm]{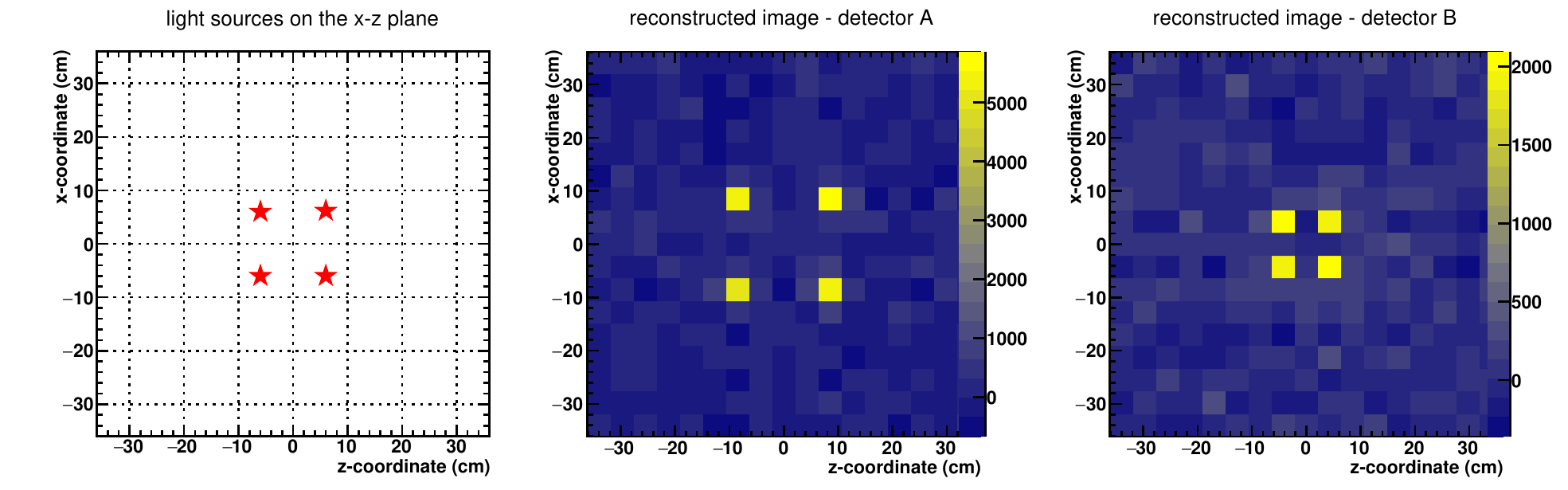}
\caption{Simulation of the 4 light-points. The left frame represents the actual sources in the $x-z$ view. The other two frames represent the reconstructed 
signal by means of two different detectors. Detector A (central frame) is closer to the sources, detector B (right frame) is farther. For detail on the image
reconstruction see Sec.~\ref{sec:4punti}.} \label{4puntipaolo}
\end{center}
\end{figure}

Simple light signals were studied to evaluate the effectiveness of the harmonic-mean method. Four point-sources have been simulated at the vertices
of a square ($12\ cm$ size) on the $x, z$ plane (Fig.~\ref{4puntipaolo}). The signal is collected by a couple of parallel detectors. The first one (A) is at
$17\ cm$ from the light-sources, the second one at $33\ cm$. The left frame in Fig.~\ref{4puntipaolo} shows the "true" positions of the light-sources. The 
signal reconstruction on the detector A is shown in the central frame. The right frame represents the signal on the opposite detector (B). From these figures, 
it is apparent that the reconstructed positions are subjected to a homogeneous scaling effect, due to the different perspective projection. Indeed the 
reconstructed image is larger for the closer detector (A) and smaller for the farther detector (B). By calculating the harmonic mean (eq.~\eqref{harmonicmean}) 
and estimating the error (eq.~\eqref{eq:error}) one gets the reconstructed coordinates $\pm (5.5 \pm 1.2)\ cm$ of the light sources. They are compatible 
within $0.42 \ \sigma$ with the actual coordinates $\pm 6.0\ cm$. Moreover, we verified that also the third coordinate $y_S$  can be correctly estimated by 
the third equation in (\ref{harmonicmean}). However, we stress that this is just a particular case as this formula for $y_S$ is not in general reliable.

For the analysis of signals more complex than 4 light-points, a preliminary procedure was implemented in order to follow the track and extract the spatial extent 
({\emph{i. e.}} the distance between the end-points) of the signal from the noise. Based on a careful investigation of the image histograms, we adopted the following 
step procedure:

\begin{itemize}
\item when a negative content is associated to a bin of the reconstructed signal, this content is substituted with its absolute value;
\item a Gaussian low-pass filter is applied to the two-dimensional distribution in order to get a better separation between noise and signal;
\item the distribution of the values of all bins is studied and fitted with a Gaussian;
\item a threshold is fixed on such a distribution:  typically we accept as signal the bins with a separation from the Gaussian 
center larger than $4\sigma$.
\end{itemize}

\noindent We know that this selection method has not been tested on a full Monte Carlo. The application of this method and the threshold choice for fully 
simulated neutrino interactions are necessary. However, we will use this preliminary signal selection as a provisional instrument to check 
the formulas found in this work.

\begin{figure}[hbp]
\begin{center}
\includegraphics[width=12.0cm]{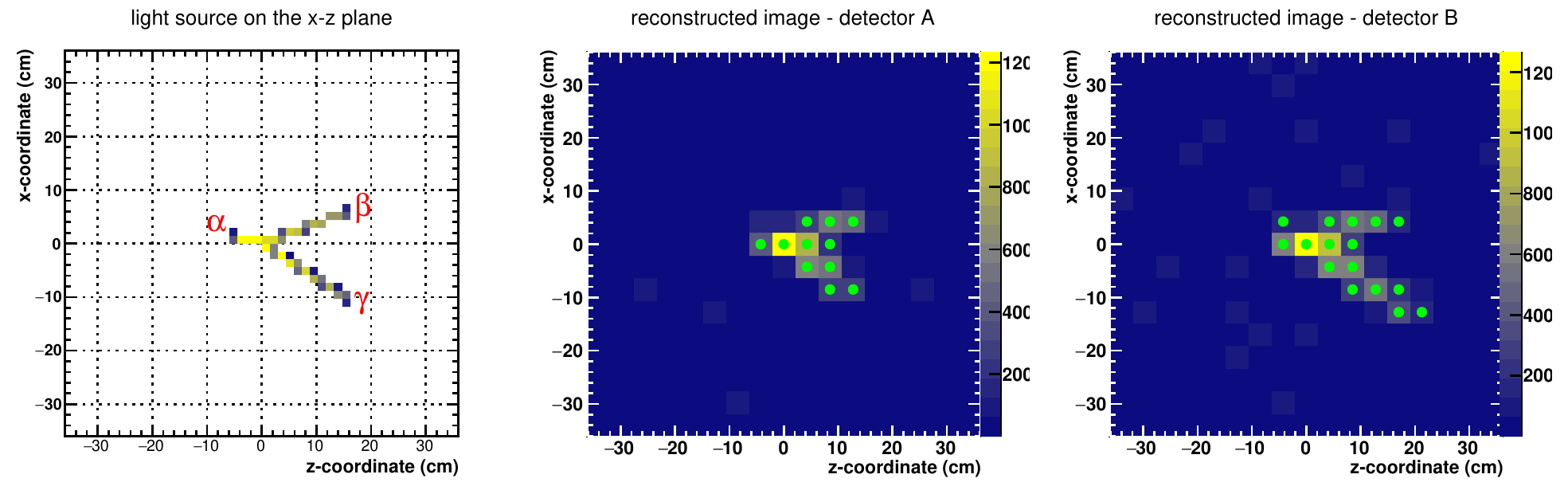}
\caption{ \emph{Left} frame - simulation of a neutrino interaction. The three tracks do not lay on the same plane. \emph{Central} and \emph{Right} frames 
show the reconstructed image in the $x-z$ view. The superimposed green dots characterize the pixels selected as signal. The position of the points $\alpha$, 
$\beta$ and $\gamma$ is estimated by means of the harmonic mean. To compare "truth" and reconstruction look at Table~\ref{tab:vertex}.} \label{Vertex}
\end{center}
\end{figure}

In Fig.~\ref{Vertex}, left frame, the tracks generated by a simulated neutrino interaction are shown in the $x-z$ view. Obviously the tracks do not lay on a 
single plane. The light signal emitted by the tracks is collected and reconstructed by two parallel masks sharing the same focal plane (Fig.~\ref{Vertex}, 
center and right). The position of the end-points ($\alpha$, $\beta$ and $\gamma$) has been reconstructed by means of the harmonic mean applied to 
the edge signal pixels. The measured coordinates are compared with the "true" ones in Table~\ref{tab:vertex} assuming the error of 
formula~\eqref{eq:error}. The agreement is fairly good. The largest discrepancy ($2\sigma$) is due to the vanishing of the reconstructed track when
the original one is too far from a detector (on the $y$-axis, perpendicular to the analysed view).

There are at least two comments concerning the presented reconstruction. First of all it does not represent a realistic simulation of a generic event that may occur in DUNE experiment, but it is a  proof of principle of how the coded mask technology may be applied in the context of the high energy Physics. On the other hand, it is clear that the chosen set up (dimensions of the mask matrix, magnification ratio in particular) leads to a resolution power, which limits the possibility to discriminate among high complex topologies. But, in principle one may scale the number of pixels in order to significantly improve the resolution, without distorting the basic concepts involved in our considerations. 

\begin{table}[h]
\caption{Reconstruction of the edge points in a neutrino-event (Fig.~\ref{Vertex}). The
reconstruction errors are calculated according to formula~\eqref{eq:error}. } \label{tab:vertex}
\centering
\begin{tabular}{lccc} \hline\noalign{\smallskip}
                                            & "truth" ($cm$) & reconstruction ($cm$) & discrepancy ($n_\sigma$) \\ \hline
$\alpha$      z-coordinate    &     -5.0            & $-4.25\pm0.87$            & 0.86 \\
$\alpha$      x-coordinate    &    +1.5            & $+2.10\pm1.23$           & 0.49 \\
$\beta$        z-coordinate    &  +16.0            & $+14.58\pm0.92$         & 1.54 \\
$\beta$        x-coordinate    &    +6.0            & $+4.25\pm0.87$           & 2.01 \\
$\gamma$   z-coordinate    &  +16.0            & $+15.95\pm1.02$         & 0.05 \\
$\gamma$   x-coordinate    &   -10.5            & $-10.21\pm0.97$          & 0.30 \\ \hline
\end{tabular}
\end{table}

\subsection{Image 3-D analysis} \label{Imanalysis}

The signal due to a linear light-track has been simulated to check the estimates of Sec.~\ref{sec:3D}. The track is described in the space by the 
following cosine directors and by an arbitrary point
\beq
M = (0.67,\ 0.67,\ -0.33) \qquad \qquad \qquad P_0 = (-3,\ 0,\ -1.5) \label{eq:truth}
\eeq
\noindent In the Cartesian notation
\beq 
y = -2\ z - 3;  \label{original}  \qquad \qquad \qquad  y = + x +3.
\eeq

\begin{figure}[hbp]
\begin{center}
\includegraphics[width=12.0cm]{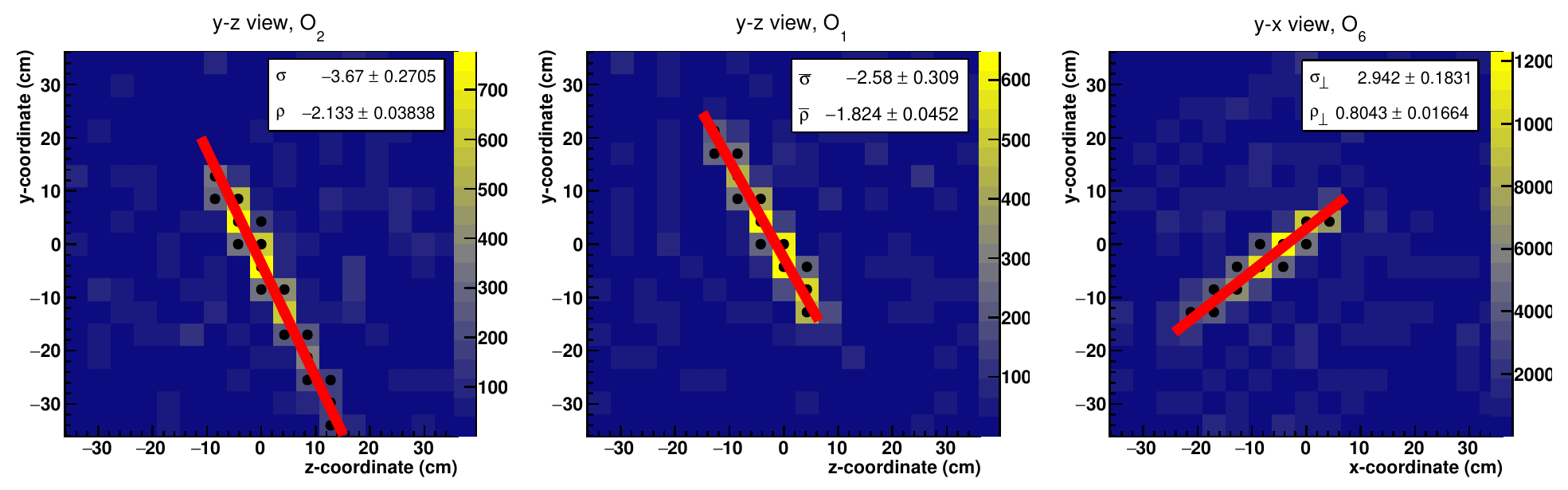}
\caption{Reconstruction of a linear track on three different detectors ($O_1$, $O_2$ and $O_6$). The linear fit is applied to the pixels (black dots) selected 
according to the preliminary algorithm of signal recognition. The fit parameters shown in the frames have been used to reconstruct the original track in 3-D. 
See Sec.~\ref{Imanalysis} for details.}\label{fig:traccia18}
\end{center}
\end{figure}

\noindent The signal has been analyzed by means of the detectors $O_1$, $O_2$ and $O_6$. The detected images are shown in Fig.~\ref{fig:traccia18}
where a linear fit is superimposed on the selected pixels (black dots). The parameters of the fit are also reported in Fig.~\ref{fig:traccia18}. The detectors 
in $O_1$ and $O_2$ are parallel and the final track in the $y-z$ view can be reconstructed according to formulas~\ref{coeffnsup}. The result is comparable
with the first eq.~\eqref{original}

\beq 
y = -1.95\ z -3.03 \label{parallel}
\eeq

Detectors $O_1$ and $O_6$ are perpendicular and allow to reconstruct the light track in the space. The fit parameters have been used in the 
formulas~\eqref{coeffnsum}, \eqref{inverted} and \eqref{p0} to get 
\beq
M = (0.72,\ 0.63,\ -0.30)  \qquad \qquad \qquad P_0 = (-3,\ 0,\ -1.5)  \label{eq:result}
\eeq

\noindent Neglecting the errors due to the pixel size, the errors on the fit parameters are enough to claim that the calculated equations are in good agreement
with the original ones. Then the results of Sec.~\ref{sec:3D} are confirmed in the frame of the single-pinhole approximation.

\section{Conclusions}
In the present work, we have reported a study concerning the application of the method of coded  masks as detectors of tracks of charged particles in scintillating media. It has been 
shown that the system actually allows for the detection of tracks over focal distances of the order of tens of centimeters. From theoretical arguments and numerical simulations, it emerges
that it is possible to implement decoding and recognition procedures for signals, even complex ones such as neutrino interactions, with a relatively limited number of channels (few 
hundreds for each SiPM array). By using a preliminary procedure of noise reduction and signal clustering, it has been shown the possibility to make measurements in agreement with the 
theoretical evaluations. If opposite and/or orthogonally arranged masks are used, the measurements can be correlated, by using simple geometrical formulas. A 3-dimensional reconstruction 
is possible, even for sources out of the focal planes. An alternative image reconstruction method is being pursued, based on the calculation of deconvolution matrices depending on the 
depth of field. This approach, which is intrinsically 3-D, might solve the problem of the limited depth of field due to the near field conditions. Other critical issues must be yet carefully studied, 
as intensity of the photon signal, detection efficiency, rejection of noise and artifacts. Therefore we are developing a full Monte Carlo in order to complete the design of a real detector. The 
implementation of such Monte Carlo is in progress in parallel with the design and construction of prototypes of the detector for the imaging of neutrino interactions in LAr. Also the feasibility 
to complement the reconstruction of such events with the timing information is under analysis.

Finally, complete validation of high complex signal reconstruction will require other deep considerations. In this regard, recently in \cite{domine}
points of interest in particle trajectories, such as 
the initial point of electromagnetic particles,
from straight
line-like tracks to branching tree-like electromagnetic
showers, have been studied by means of Neural Network, significantly improving
the efficiency of finding candidate interaction vertices, and
hence candidate neutrino interactions, which may be used in  high-level physics inference, for instance in the context of the DUNE and SBND experiments. Those developments merit our attention in further dedicated efforts.

\section{Acknowledgments}
The present research  is supported by the Italian \emph{Ministero dell'Universit\`a e della Ricerca} (PRIN 2017KC8WMB) and by the \emph{Istituto Nazionale Fisica Nucleare}
(experiments NU@FNAL and MMNLP).

%%%%%%%%%%%%%%%%%%%
\end{document}